\documentclass[%
 reprint,
 superscriptaddress,
 amsmath,amssymb,
 aps,
 pra,
]{revtex4-2}

\usepackage{graphicx}
\usepackage{dcolumn}
\usepackage{bm}

\usepackage[utf8]{inputenc}
\usepackage[version=4]{mhchem}
\usepackage{enumitem}
\usepackage{csquotes}
\usepackage{nicefrac}
\usepackage{mathtools}
\usepackage{xcolor}
\usepackage{siunitx}
\usepackage{url}
\usepackage[colorlinks=true, 
            linkcolor=black, 
            urlcolor=blue, 
            citecolor=black]{hyperref}

\renewcommand{\selectlanguage}[1]{} 

\newcommand{\textcomment}[1]{}

\newcommand{\heone}{\ce{^{4}He^{1+}}}
\newcommand{\hetwo}{\ce{^{4}He^{2+}}}
\newcommand{\hethreeone}{\ce{^{3}He^{1+}}}
\newcommand{\cthree}{\ce{^{12}C^{3+}}}
\newcommand{\cfour}{\ce{^{12}C^{4+}}}
\newcommand{\csix}{\ce{^{12}C^{6+}}}
\newcommand{\ofour}{\ce{^{16}O^{4+}}}

\newcommand{\oeight}{\ce{^{16}O^{8+}}}
\newcommand{\nseven}{\ce{^{14}N^{7+}}}
\newcommand{\hthree}{\ce{H_{3}^{1+}}}
\newcommand{\hone}{\ce{^{1}H^{1+}}}

\newcommand{\ebgmedaustrongmbh}{EBG MedAustron GmbH Marie-Curie Straße 5, 2700 Wiener Neustadt, Austria}
\newcommand{\ati}{Atominstitut, TU Wien, Stadionallee 2, 1020 Vienna, Austria}
\newcommand{\hephy}{Institute of High Energy Physics of the Austrian Academy of Sciences, Nikolsdorfer Gasse 18, 1050 Vienna, Austria}
\newcommand{\meduni}{Medical University of Vienna, Spitalgasse 23, 1090 Vienna, Austria}
\newcommand{\gsi}{GSI Helmholtzzentrum für Schwerionenforschung GmbH, Planckstraße 1, 64278 Darmstadt, Germany}

\begin{document}

\preprint{APS/123-QED}

\title{\texorpdfstring{A double multi-turn injection scheme for generating \\ mixed helium and carbon ion beams at medical synchrotron facilities}{A double multi-turn injection scheme for generating mixed helium and carbon ion beams at medical synchrotron facilities}}

\author{Matthias Kausel}
    \email{Contact author: matthias.kausel@medaustron.at}
    \affiliation{\ebgmedaustrongmbh}
    \affiliation{\ati}

\author{Claus Schmitzer}
    \affiliation{\ebgmedaustrongmbh}%

\author{Andreas Gsponer}
    \affiliation{\hephy}
    \affiliation{\ati}

\author{Markus Wolf}
    \affiliation{\ebgmedaustrongmbh}%

\author{Hermann Fuchs}
    \affiliation{\meduni}

\author{Felix Ulrich-Pur}
    \affiliation{\gsi}

\author{Thomas Bergauer}
    \affiliation{\hephy}

\author{Albert Hirtl}
    \affiliation{\ati}

\author{Nadia Gambino}
    \affiliation{\ebgmedaustrongmbh}%

\author{Elisabeth Renner}
    \affiliation{\ati}


\begin{abstract}
The low relative charge-to-mass ratio offset of 0.065\,\% between fully ionized helium-4 and carbon-12 ions enables simultaneous acceleration in hadron therapy synchrotrons. At the same energy per mass, helium ions exhibit a stopping range approximately three times greater than carbon ions. They can therefore be exploited for online range verification downstream of the patient during carbon ion beam irradiation. One possibility for creating this mixed beam is accelerating the two ion species sequentially through the LINAC and subsequently \enquote{mixing} them at injection energy in the synchrotron with a double multi-turn injection scheme.

This work reports the first successful generation, acceleration, and extraction of a mixed helium and carbon ion beam using this double multi-turn injection scheme, which was achieved at the MedAustron therapy accelerator in Austria. A description of the double multi-turn injection scheme, particle tracking simulations, and details on the implementation at the MedAustron accelerator facility are presented and discussed. Finally, measurements of the mixed beam at delivery in the irradiation room using a radiochromic film and a low-gain avalanche diode (LGAD) detector are presented.
\end{abstract}

\maketitle

\section{Introduction}
\begin{figure}
    \centering
    \includegraphics{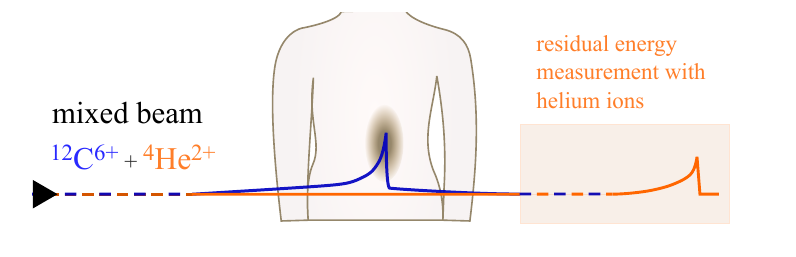}
    \caption{Schematic mixed beam irradiation. The carbon beam is used for tumor treatment while the residual helium energy is measured downstream of the patient for diagnostic purposes.}
    \label{fig:schematicMixedBeamIrrad}
\end{figure}

\noindent In recent years, a new concept of reducing range uncertainties via irradiation with mixed helium and carbon ion beams was proposed and assessed \cite{graeff_helium_2018, mazzucconi_mixed_2018, volz_experimental_2020, hardt_potential_2024}. Exploiting the approximately three times higher range of helium ions in matter compared to carbon ions at the same energy per mass holds the potential for simultaneous carbon ion tumor treatment and helium radiography downstream of the patient, as illustrated in Fig.~\ref{fig:schematicMixedBeamIrrad}. The information obtained by the helium energy loss inside the patient can be used to reconstruct the range of carbon within the patient during the treatment. It was estimated that the overall dose increase is less than 1\,\% for a beam mixture of 90\,\% carbon and 10\,\% helium ions in the treatment beam \cite{graeff_helium_2018, mazzucconi_mixed_2018}. This dose increase is considered acceptable as it is within the current treatment tolerances \cite{mazzucconi_mixed_2018}.

Until recently, studies on mixed beam irradiation focused primarily on simulations \cite{graeff_helium_2018, mazzucconi_mixed_2018, hardt_potential_2024} and experiments using sequential irradiation with carbon and helium ions or protons \cite{mazzucconi_mixed_2018, volz_experimental_2020}. A significant milestone was reached when the first mixed \heone\ and \cthree\ beam was delivered at the GSI Helmholtz Center for Heavy Ion Research in late 2023 \cite{galonska_first_2024, ondreka_slow_2024}. This beam was produced in a single ion source and subsequently accelerated and slow extracted at 225\,MeV/u. While it is possible to accelerate \heone\ and \cfour\ beams at heavy ion research facilities, such as e.g.\ at GSI, their charge-to-mass ratio of approximately $\nicefrac{1}{4}$ is not within the injector design specifications for state-of-the-art ion therapy accelerators, such as HIT~\cite{haberer_heidelberg_2004, volz_experimental_2020}, CNAO~\cite{amaldi_cnao-italian_2004} or MedAustron~\cite{benedikt_medaustron_2016}.

As will be detailed in Section~\ref{ch:evaluating_options}, also for other conceivable combinations of helium and carbon ions with similar charge-to-mass ratio, simultaneous generation and injection from a single ion source is challenging, given the current infrastructure at these facilities. Therefore, an alternative method of creating the ion mix via sequential injection of helium and carbon ions into the synchrotron is being investigated. This approach, implemented as a double multi-turn injection scheme at the MedAustron accelerator facility, enabled the first delivery of a mixed \hetwo\ and \csix\ beam in a medical synchrotron facility. It could, in principle, also be employed at other medical synchrotron facilities, assuming the respective control system allows for such a sequential injection scheme.

The first part of this work describes the sequential injection scheme and its necessity for mixed helium and carbon ion beam generation at the MedAustron accelerator facility. The second part elaborates on the experimental implementation and presents the results of the first 262.3\,MeV/u mixed \hetwo\ and \csix\ beam extraction into the non-clinical irradiation room at MedAustron, including its detection using a radiochromic film and silicon low-gain avalanche diode (LGAD) detector measurement.

\section{\texorpdfstring{The MedAustron\\ Accelerator Facility}{The MedAustron Accelerator Facility}}
\label{ch:MedAustronAcceleratorFacility}
\begin{figure*}
    \centering
    \includegraphics[width=\linewidth]{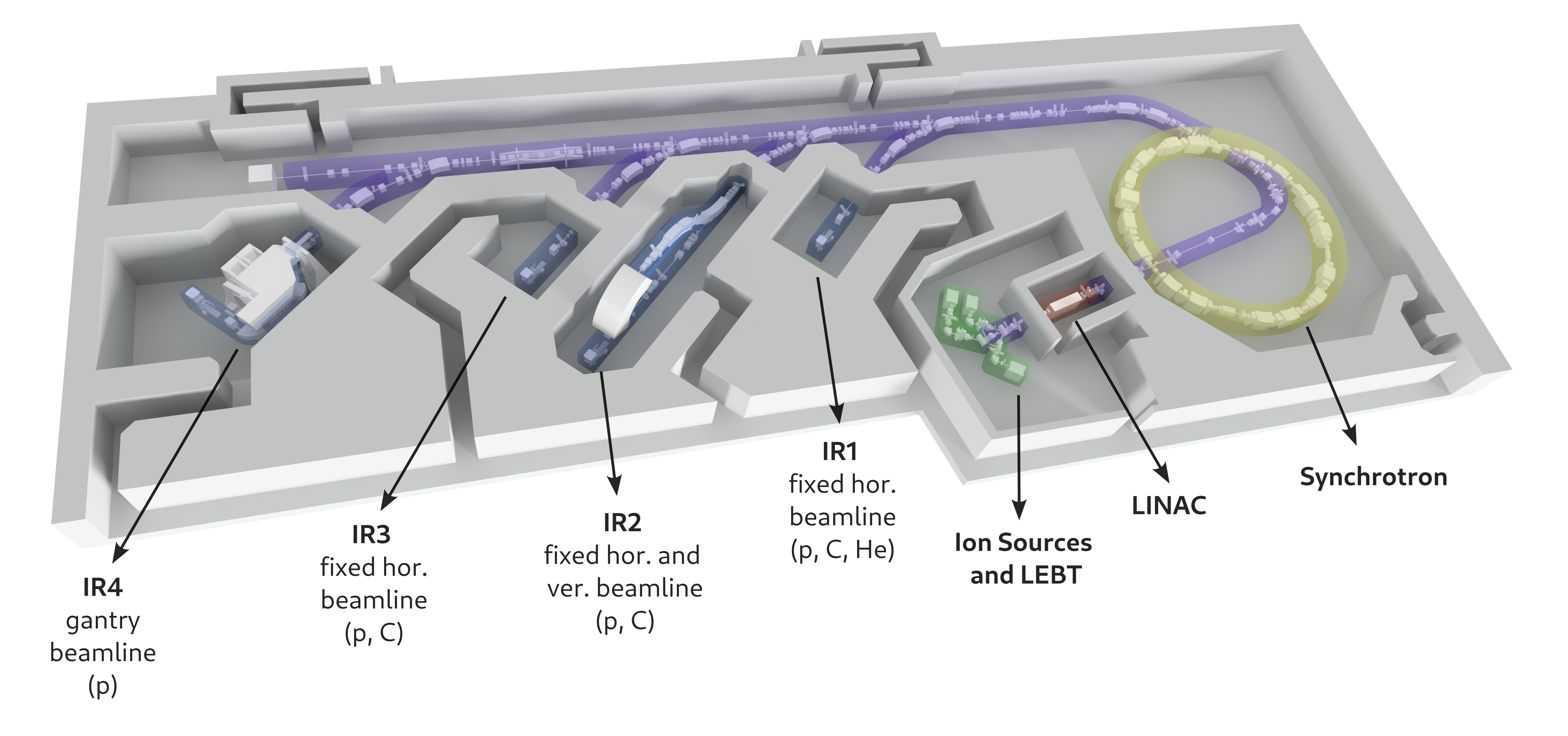}
    \caption{Layout of the MedAustron accelerator adapted from~\cite{pivi_commissioning_2024}. The availability of proton (p), carbon (C), or helium ions (He) is indicated for each irradiation room.}
    \label{fig:MedAustronLayout}
\end{figure*}

\noindent The MedAustron accelerator facility in Wiener Neustadt, Austria, features a medical synchrotron with 78\,m circumference for ion radiotherapy and research \cite{benedikt_medaustron_2016} based on the proton-ion medical machine study (PIMMS) design \cite{badano_proton-ion_2000, bryant_proton-ion_2000}. With the availability of \qtyrange{120}{402.8}{MeV/u} clinical \csix\ \cite{pivi_status_2019} and \qtyrange{39.8}{402.8}{MeV/u} \hetwo\ beams for non-clinical research \cite{gambino_status_2024}, MedAustron provides the necessary prerequisites for investigating the proposed sequential injection scheme. The facility, which is depicted in Fig.~\ref{fig:MedAustronLayout}, features three identical electron-cyclotron resonance (ECR) ion sources producing \hthree, \cfour\ and \hetwo, respectively. The beams are extracted at 8\,keV/u and injected into the linear accelerator (LINAC) section, which consists of a radio-frequency quadrupole (RFQ) and an interdigital H-mode (IH) drift tube structure. After acceleration to around 7\,MeV/u in the LINAC, the beam passes a stripping foil which strips \hthree\ to \hone\ and \cfour\ to \csix, while \hetwo, as a fully ionized particle, is not affected. The ions are injected into the synchrotron, captured, and accelerated to the desired energy. The installed synchrotron radio frequency (RF) system currently only provides single harmonic operation, i.e.\ it only provides one single frequency. After the acceleration, the ions are slow extracted via a third-order resonance extraction driven by a betatron core \cite{badano_characteristics_1997}. Besides the betatron core, alternative excitation methods, such as phase displacement or RF knock-out schemes, are currently being investigated \cite{renner_towards_2024, kuhteubl_slow_2024}. The extracted beam is steered into one of four irradiation rooms IR1-IR4. Three of these rooms (IR2-IR4) are dedicated to clinical irradiation, while IR1 is used for non-clinical research \cite{schreiner_medaustron_2019}.

\subsection{Synchrotron injection system}
\label{ch:injectionSystem}
\noindent At MedAustron, the beam is injected into the synchrotron via a coasting beam multi-turn injection. The injection septum is located in a dispersion-free region with an aperture of 41\,mm between the on-momentum beam orbit and the septum blade. Two kicker magnets create an orbit bump, which moves the beam trajectory toward the septum. The injection bump decays linearly in approximately 80\,µs and ions are injected within an injection window of around 30\,µs, corresponding to approximately 14 turns.

In nominal operation with mono-isotopic beams, the synchrotron optics are configured to inject the ions with a relative momentum offset $\Delta p{/}p\approx-0.0024$. Considering the maximum dispersion of $D_{x,\text{max}}\approx-8.4$\,m, this corresponds to a maximum dispersive offset of around 20\,mm toward the inner wall of the synchrotron vacuum chamber.

All ion species are injected with kinetic energies $E{/}m\approx7$\,MeV/u. However, different configurations in the LINAC, required due to the different charge-to-mass ratios before stripping, result in slight discrepancies in the injection energy for the different ion species. The commissioned \hetwo\ beam is injected into the synchrotron at around 7.07\,MeV/u, whereas the commissioned \csix\ ions are injected at approximately 6.97\,MeV/u. The synchrotron magnets are configured accordingly to ensure the desired dispersive offset as well as minimal losses during the capture process for each ion type. In the injector both carbon and helium ions feature horizontal and vertical RMS normalized emittances $\epsilon_{\text{n},x,y}^\text{inj} \approx 2\cdot10^{-7}$\,m\,rad and an RMS relative momentum spread $\sigma_{\nicefrac{\Delta p}{p}}\approx7\cdot10^{-4}$. To achieve comparable beam emittances in the horizontal and vertical planes after the multi-turn injection, the vertical beta function of the injected beam is deliberately mismatched relative to the beta function at the injection septum. This intentional mismatch leads to dilution of the vertical emittance. After the multi-turn injection, the accumulated beam features normalized RMS emittances of $\epsilon_{\text{n},x,y} \approx 7\cdot10^{-7}$\,m\,rad. The nominal injection intensity is around $1.7\cdot10^{8}$\,ions/turn (75\,µA) for carbon and $1.2\cdot10^{9}$\,ions/turn (180\,µA) for helium.

\subsection{\texorpdfstring{Evaluating options for generating mixed beams \\ in the ion sources at MedAustron}{Evaluating options for generating mixed beams in the ion sources at MedAustron}}
\label{ch:evaluating_options}

\noindent In the following, we evaluate the options for generating a mixed helium and carbon ion beam directly in an ion source and discuss why this is not feasible with the current infrastructure available at MedAustron.

In the synchrotron, the two ion species must have a similar beam rigidity $B\rho$ to maintain a horizontal offset between their respective dispersive orbits
\begin{equation}\label{eq:dispoffset}
    \Delta x = D \cdot \frac{\Delta\left(B\rho\right)}{B\rho}
\end{equation} 
which is compatible with the aperture restrictions in the lattice.
As elaborated in Appendix~\ref{ch:appendixRequirements}, in synchrotrons with one single-harmonic RF system such as the MedAustron accelerator, only ion pairs with nearly identical charge-to-mass ratios can fulfill this condition during acceleration as the revolution frequencies need to be similar and therefore $\Delta (\beta\gamma)/(\beta\gamma)\ll1$. Consequently, the following discussion focuses on ion mixtures with charge-to-mass ratio differences $<1\,\%$. In principle, there would be three options for generating such an ion mix in a single ion source.

\noindent\paragraph*{\textbf{\heone\ and \cthree}:}
The production of a mixed \heone\ and \cthree\ beam within a single ion source, as done for the measurements at the GSI Helmholtz Center for Heavy Ion Research \cite{galonska_first_2024}, would theoretically be feasible with the sources installed at the MedAustron facility. After pre-acceleration in the LINAC, the ion pair could be stripped to \hetwo\ and \csix\ for injection into the synchrotron and subsequent acceleration. However, the LINAC design specifications only allow for the acceleration of ions with a charge-to-mass ratio larger than \nicefrac{1}{3}. Consequently, a \heone\ and \cthree\ ion beam, featuring a charge-to-mass ratio of approximately $\nicefrac{1}{4}$, cannot be pre-accelerated and injected into the synchrotron. A stripping of \heone\ and \cthree\ to \hetwo\ and \csix\ before pre-acceleration in the LINAC at $E{/}m=8$\,keV/u is not feasible due to the reduced stripping efficiency and short penetration depths of the ions at low energies~\cite{bryant_proton-ion_2000}.

\noindent\paragraph*{\textbf{\hetwo\ and \csix:}}
The charge-to-mass ratios in a mixed \hetwo\ and \csix\ beam would fall within the LINAC's design specifications, allowing them to be simultaneously injected and accelerated. However, for the Pantechnik Supernanogan ion sources installed at MedAustron, the yield of \csix\ is approximately two orders of magnitude lower than that of \cfour\ \cite{noauthor_pantechnik_2013}. At these intensities, the treatment times would be significantly prolonged, which renders the Supernanogan sources unsuitable for clinical \csix\ production. Newer generation ion sources, such as e.g.\ the AISHa ion source \cite{castro_aisha_2022}, could deliver higher extracted \csix\ intensities. 
 
\noindent\paragraph*{\textbf{\hethreeone\ and \cfour:}}
Alternatively, also a mixed \hethreeone\ and \cfour\ beam could be generated using the existing ion sources at MedAustron by using \ce{^3He} instead of \ce{^4He} as source gas. This option is impractical as no further stripping after pre-acceleration in the LINAC is possible. The smaller charge-to-mass ratio of $\nicefrac{1}{3}$ reduces the achievable energy in the synchrotron, and as a consequence, the available treatment depth of the carbon ions. Moreover, the offset in charge-to-mass ratio $\Delta\left(\nicefrac{q}{m}\right)/\left(\nicefrac{q}{m}\right)\approx-5.3\cdot10^{-3}$ is approximately one order of magnitude larger compared to the charge-to-mass ratio offset for \hetwo\ and \csix\ $\Delta\left(\nicefrac{q}{m}\right)/\left(\nicefrac{q}{m}\right)\approx-6.5\cdot10^{-4}$. This larger offset in charge-to-mass ratio for a mixed \hethreeone\ and \cfour\ beam is problematic, as during the capture and acceleration with a single-harmonic RF system, it implies a larger rigidity offset according to equation~\ref{eq:appendix_rigidity_offset} in Appendix~\ref{ch:appendixFrequencyOffset}. Figure~\ref{fig:envelope_different_ion_pairs} illustrates the respective dispersive beam orbits and RMS envelopes at injection energy $E/m\approx7$\,MeV/u in the MedAustron synchrotron, considering this rigidity offset. It is immediately apparent that a mixed \hethreeone\ and \cfour\ beam captured with a single RF frequency would be lost at the aperture.

\begin{figure}
    \centering
    \includegraphics[width=\linewidth]{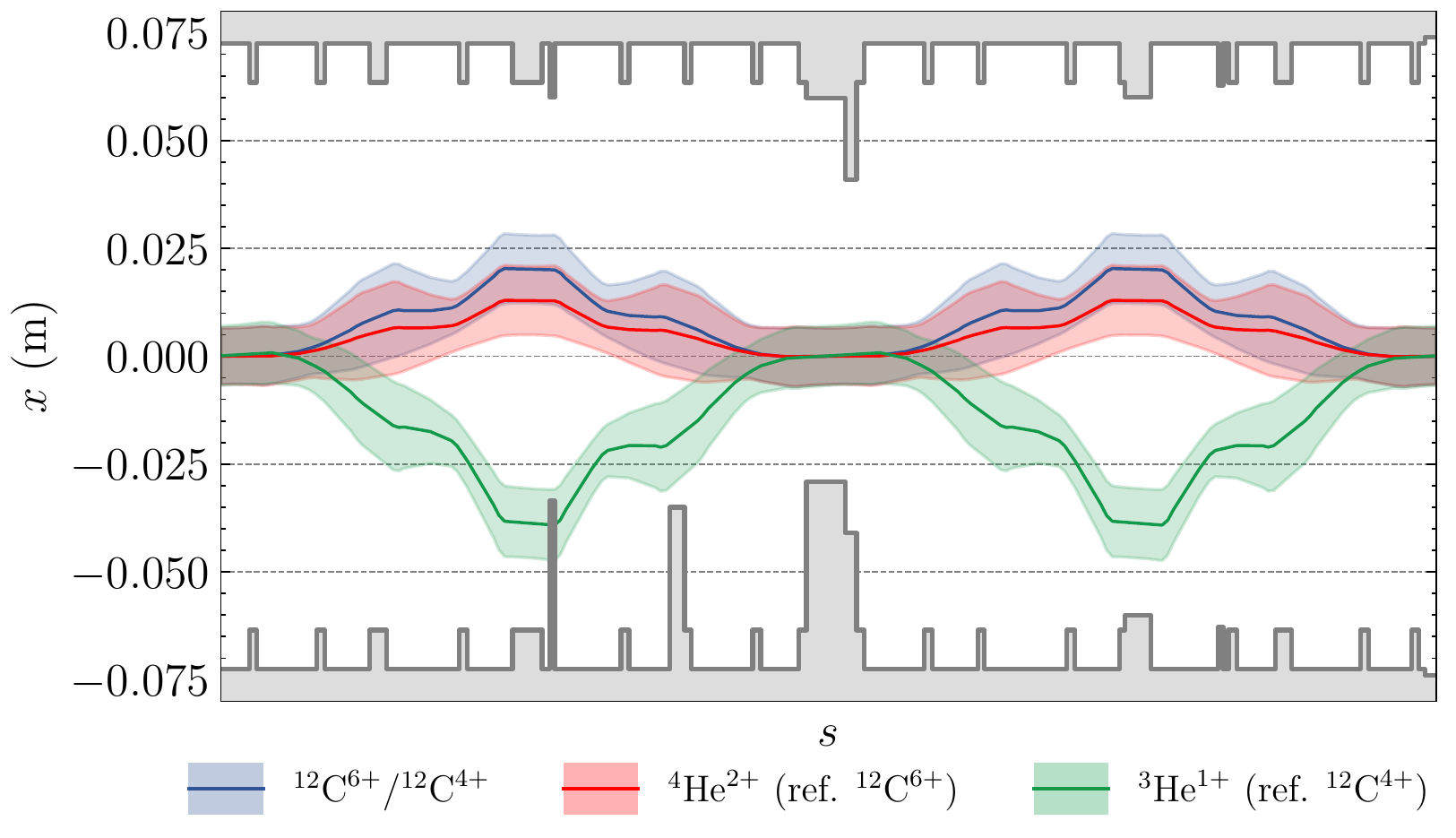}
    \caption{Dispersive beam orbits $x(s)=D_x(s)\frac{\Delta B\rho}{B\rho}$ and RMS envelopes $\sigma_x=\sqrt{\frac{\epsilon_{n,x}\beta_x(s)}{\beta\gamma}+\left(D_x(s)\sigma_{\nicefrac{\Delta B\rho}{B\rho}}\right)^2}$ for different ion pairs after the multi-turn injection at $E/m\approx7$\,MeV/u according to the parameters presented in Section~\ref{ch:injectionSystem}. The offset in $B\rho$ was chosen such that the ions exhibit the same revolution frequency (single-harmonic RF operation).}
    \label{fig:envelope_different_ion_pairs}
    \vspace{-10 pt}
\end{figure}

Given the constraints, it can be concluded that without significant machine upgrades, there is no satisfactory method of creating and accelerating a mixed helium and carbon ion beam from a single ion source at the MedAustron accelerator facility. Therefore, in the mid-term, the sequential injection of helium and carbon from the respective ion sources into the synchrotron is the only option for producing the mixed beam at MedAustron and other PIMMS-like~\cite{badano_proton-ion_2000, bryant_proton-ion_2000} medical synchrotron facilities.
\vspace{-20 pt}
\section{\texorpdfstring{The double multi-turn\\injection scheme}{The double multi-turn injection scheme}}
\label{ch:DMTIScheme}
\begin{figure}
    \centering
    \includegraphics[width=\linewidth]{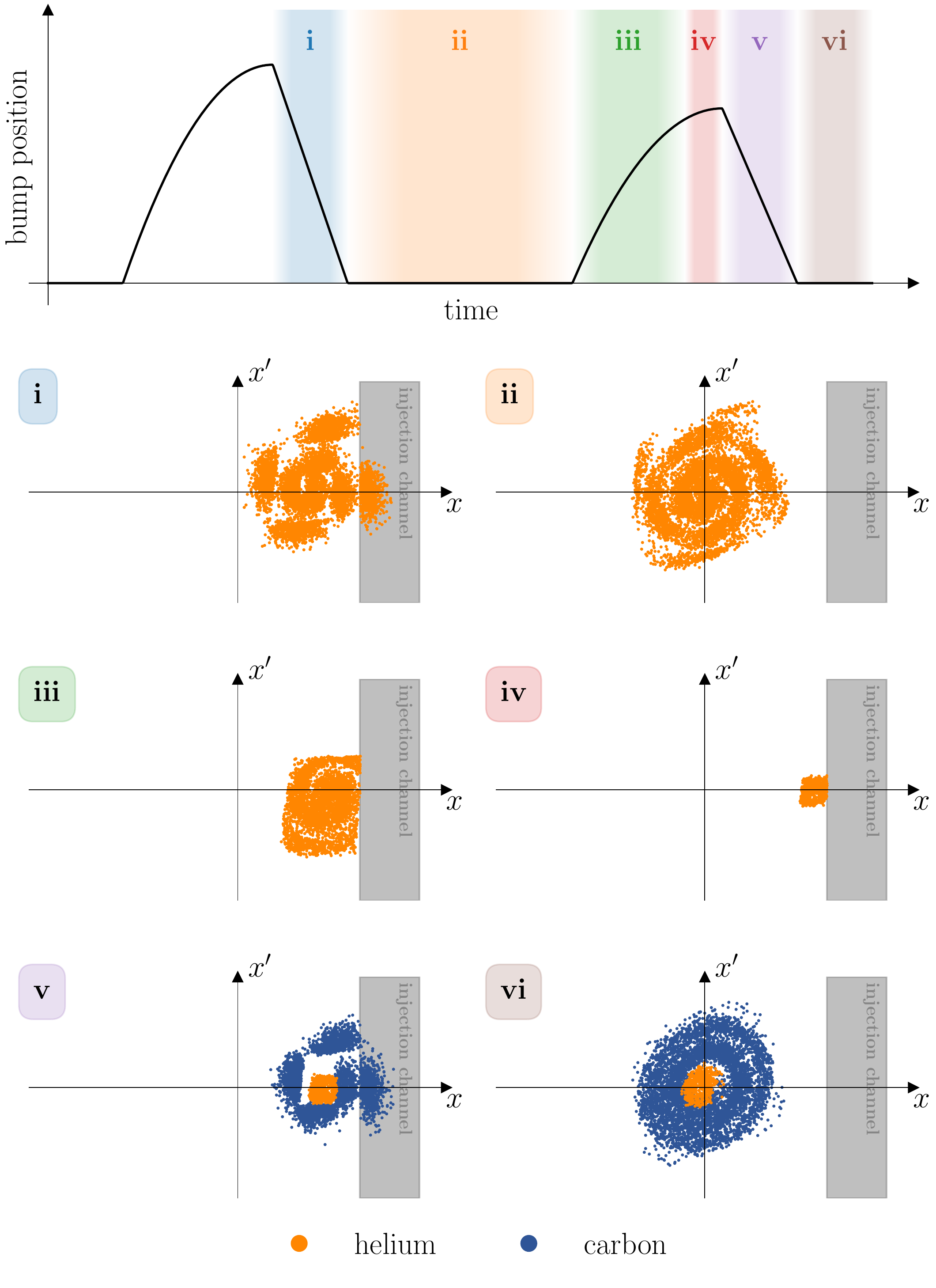}
    \caption{Schematic illustration of the double multi-turn injection scheme. Top: Time evolution of the orbit bump amplitude at the injection septum. Bottom: Horizontal phase space distributions during the double multi-turn injection.}
    \label{fig:doubleBumpSchematics}
\end{figure}

\noindent In the double multi-turn injection scheme, helium and carbon ions, generated in separate ion sources, are sequentially injected into the synchrotron to be subsequently accelerated simultaneously. This scheme, which the authors already described in~\cite{kausel_double_2024}, requires pulsing the injector twice while ramping the synchrotron only once. Figure~\ref{fig:doubleBumpSchematics} illustrates this proposal and serves as a reference for the following description.

\begin{enumerate}[itemsep=0pt, parsep=2pt, label=\roman*.]
    \item In the first injection cycle, helium is injected into the synchrotron via a multi-turn injection with nominal injection bump amplitude and decay. 
    \item As already described in Section~\ref{ch:MedAustronAcceleratorFacility}, the LINAC accelerates \cfour\ and \hetwo\ ions with charge-to-mass ratios of \nicefrac{1}{3} and \nicefrac{1}{2}, respectively. Consequently, a reconfiguration of the injector is necessary to accommodate the different charge-to-mass ratio of the carbon ions upstream of the stripping foil. During this reconfiguration, the helium beam is maintained as a coasting beam at flat bottom in the synchrotron. 
    \item Once the injector is set up for carbon, another injection cycle is triggered. This also includes ramping the injection bump in the synchrotron a second time, which causes the majority of the already circulating helium beam to be scraped at the electrostatic injection septum blade. 
    \item The amount of helium losses depends on the maximum amplitude of this second injection bump. To keep helium ions in the synchrotron, it must be reduced compared to nominal operation.
    \item During the carbon injection, a clinically acceptable carbon intensity and emittance \cite{badano_proton-ion_2000, bryant_proton-ion_2000} despite injecting with a reduced injection bump amplitude needs to be assured. A compromise of the overall carbon beam quality and helium content has to be established by choosing the bump amplitude accordingly.
    \item After the double multi-turn injection, the horizontal phase space distribution of the mixed beam features a helium core, with carbon being accumulated at larger actions.
\end{enumerate}

\noindent The resulting intensities and horizontal beam distributions of carbon and helium ions are significantly affected by the injection bump amplitude ramp and decay characteristics.
As indicated in Fig.~\ref{fig:doubleBumpSchematics}, the orbit bump ramp approximately follows a quadratic function. The bump amplitude decay is linear, with programmable maximum amplitude and decay rate. Changing the characteristics of the ramp is not possible with the hardware available at MedAustron. Therefore, the following analytical considerations and simulations assume a quadratic amplitude increase and linear decay.

\begin{figure}
    \centering
    \includegraphics[width=\linewidth]{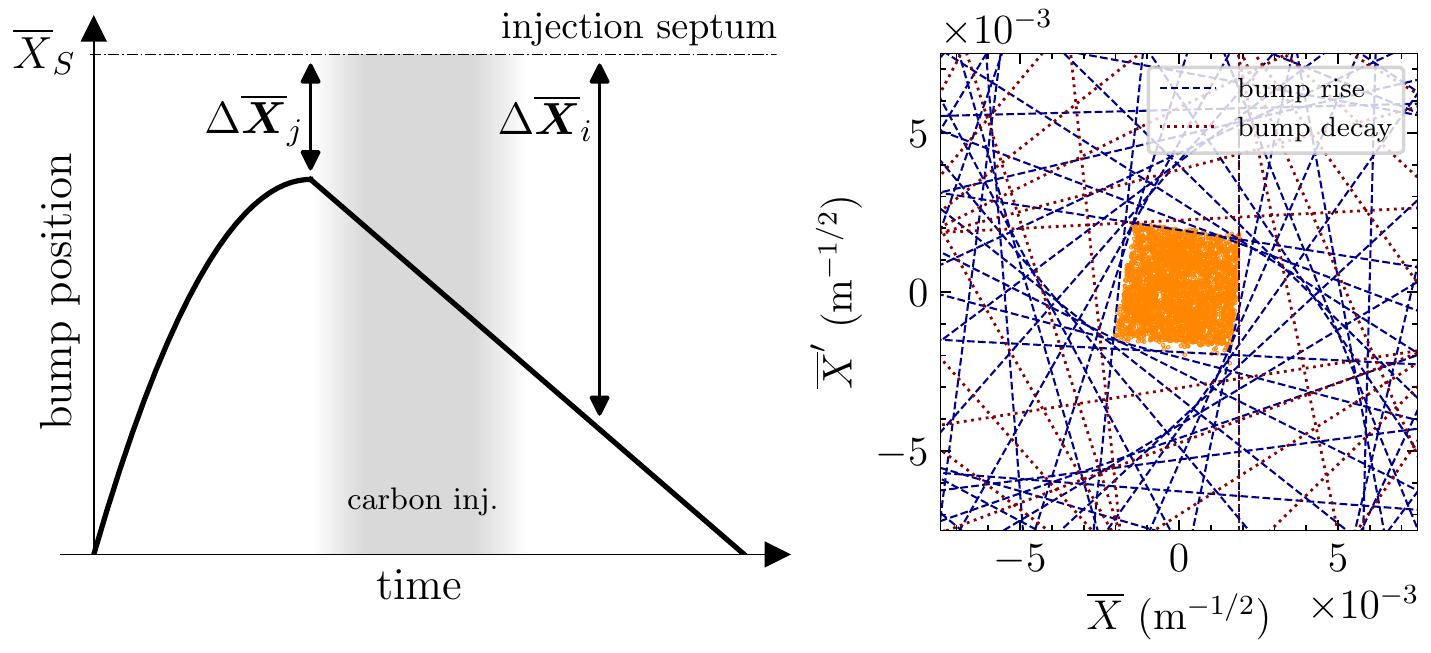}
    \includegraphics[width=\linewidth]{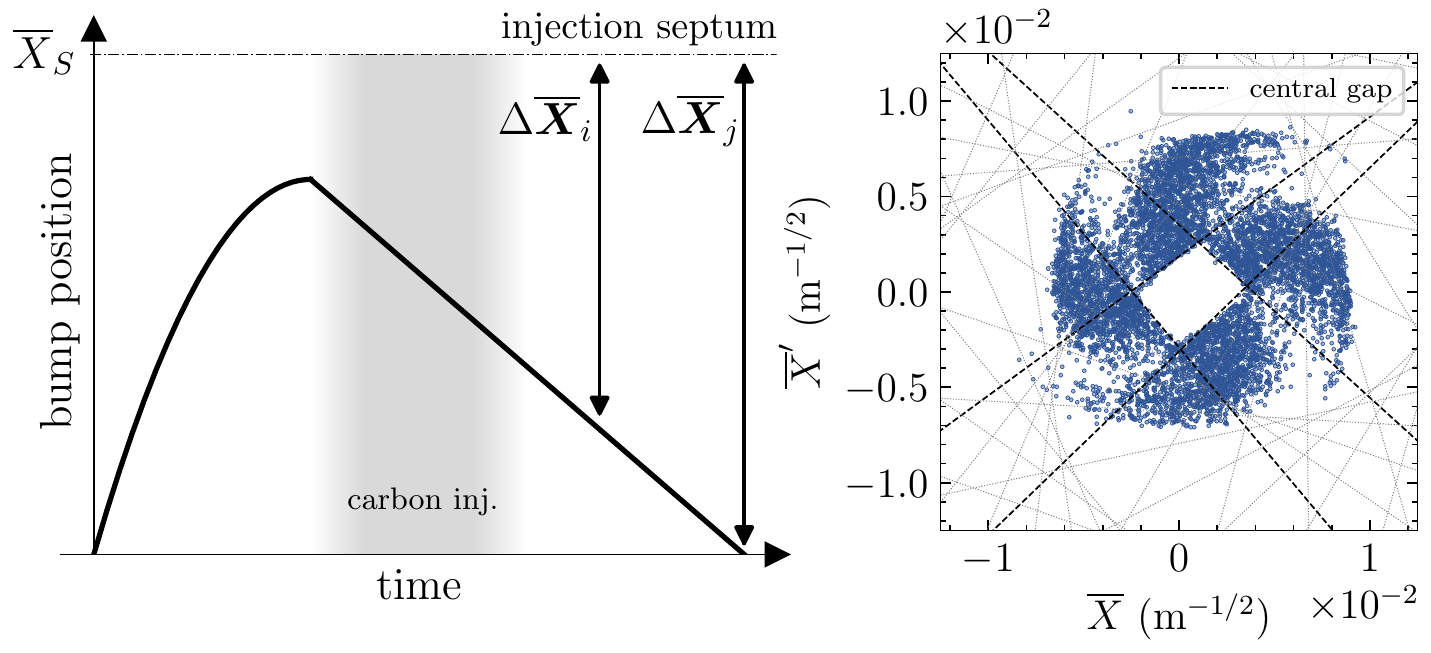}
    \caption{Analytic approximations and comparative simulations of the final helium (top) and carbon (bottom) distribution in normalized phase space $(\overline{X}, \overline{X}')$. For the helium ions, the distribution is observed at the turn at which the second (carbon) injection bump amplitude reaches its maximum. For the carbon ions, the observation turn is chosen directly after the second injection bump has fully decayed. Left: Evolution of the second injection bump, indicating the varying aperture constraints $\Delta \overline{\bm{X}}$ as distances of beam center to injection septum blade for an arbitrary turn $i$ and the observation turn $j$. Right: Normalized phase space plots displaying the analytically estimated acceptance and simulated beam distributions after the second injection bump. The total area occupied by the carbon ions is not primarily defined by the losses but is limited by the injection window (gray-shaded area). Note the different axis scaling in the top and bottom plots.}
    \label{fig:distributionAnalysis}
\end{figure}

The resulting horizontal distributions of the helium and carbon ions in normalized phase space $(\overline{X}, \overline{X}')$ can be intuitively understood and visualized by transforming the aperture limitations of the injection septum turn-by-turn using the synchrotron's linear one-turn transfer matrix. The varying acceptance during each turn of the bump ramp and decay is indicated by a respective constraint line in Fig.~\ref{fig:distributionAnalysis} (top right) and Fig.~\ref{fig:distributionAnalysis} (bottom right), which also display 
\textit{Xsuite}~\cite{iadarola_xsuite_2024} particle tracking simulations of a mono-energetic on-momentum helium and carbon ion beam for comparison. The transformation of the aperture constraint at an arbitrary turn~$i$ to the observation turn~$j$ is derived in Appendix~\ref{ch:appendixDetailsAnalysisHorizontalPhaseSpace}.
It is evident in Fig.~\ref{fig:distributionAnalysis}~(top), that the surviving helium distribution in first order, i.e.\ when neglecting filamentation, is primarily determined by the constraints associated with the last four turns of the bump rise (blue lines) rather than the decay (red lines). Figure~\ref{fig:distributionAnalysis}~(bottom) further shows that the central gap in the carbon distribution is well-defined by the constraints associated with the first four turns of the orbit bump decay. The detuning of particles with larger actions can be attributed to the settings of the chromaticity correction sextupoles. Nonetheless, for a reasonably small gap size, as expected for the desired mixing ratio, the central gap is reproduced well by the innermost constraints. It is worth noting that the carbon gap is larger than the remaining helium core. This is because the carbon gap size is solely determined by the linear bump decay, which moves the beam away from the septum faster than the quadratic rise approaches it. The larger carbon gap size is also visible in Fig.~\ref{fig:doubleBumpSchematics} step v.
 
The presented visualization of the surviving helium emittance and gap in the carbon distribution provides an intuitive understanding of the horizontal phase space features produced by the sequential multi-turn injection in first order. The dependency on the functional form of the injection bump becomes apparent, which renders the method especially useful for identifying potential future optimizations to the bump rise and decay characteristics. However, for modeling the real beam including properties such as the off-momentum operation, a finite momentum spread, or injection energy offsets, particle tracking simulations need to be performed.

\begin{figure*}[t]
    \centering
    \includegraphics[width=0.95\linewidth]{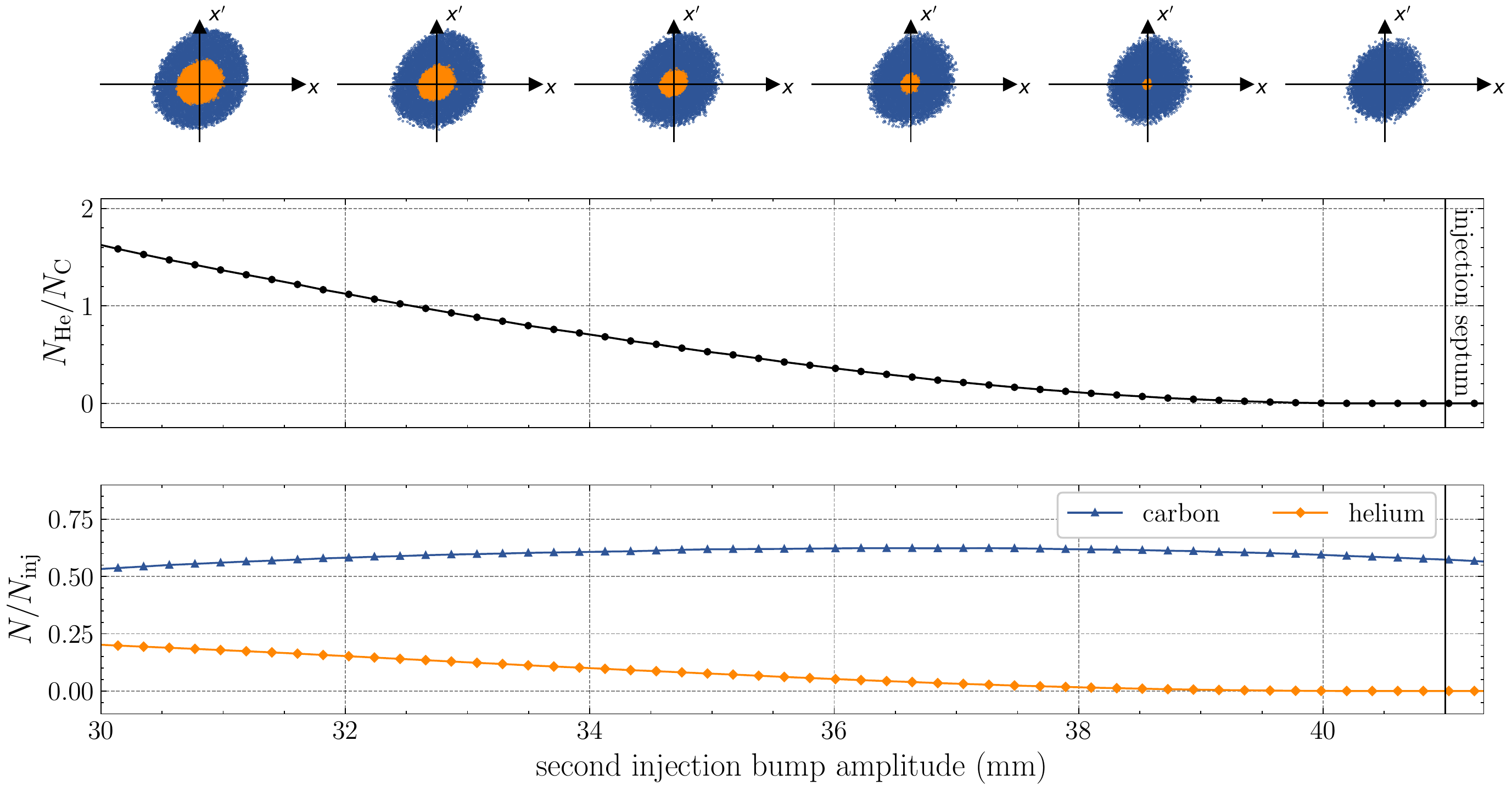}
    \caption{Impact of the second injection bump amplitude on the simulated injection efficiency of both ion species. The injection septum is located at 41\,mm from the synchrotron reference trajectory. Top: Illustration of simulated horizontal phase space distributions. Middle: Ratio of helium-to-carbon intensity after the second orbit bump, before capture and acceleration. Bottom: Injection efficiency per ion species.}
    \label{fig:particleTrackingSimulations}
    \vspace{-5pt}
\end{figure*}

\section{Particle Tracking Simulations}
\label{ch:particleTrackingSimulation}

\noindent To characterize the sensitivity of the mixed beam distribution to different amplitudes of the second injection bump \textit{Xsuite}~\cite{iadarola_xsuite_2024} particle tracking simulations were performed. In the presented proof-of-principle simulations, the initial bump amplitude was varied, while keeping the bump rise and decay times constant.

The simulation parameters are summarized in Table~\ref{tab:simulationInput}. The optical and injection steering parameters were set according to the design specifications of a PIMMS-like synchrotron \cite{badano_proton-ion_2000, bryant_proton-ion_2000}. The relative momentum spread and beam emittance at injection were chosen according to the values presented in Section~\ref{ch:injectionSystem}. Furthermore, as also described in Section~\ref{ch:injectionSystem}, the operational helium injection energy $E/m=7.07$\,MeV/u is larger than the 6.97\,MeV/u for carbon ions. To facilitate the simultaneous transport of both ion species in the synchrotron, the injection energy of helium is reduced to 7.03\,MeV/u, as will be detailed in Section~\ref{ch:injectionEnergyAdapation}. This energy reduction leads to a decrease in helium injection intensity, which was assumed to be around $6.6\cdot10^{8}$\,ions/turn (100\,µA). For these proof-of-principle simulations space charge effects are neglected, considering the \csix\ intensities of around $~10^9$ ions, the respective \hetwo\ contributions, and the space charge estimations performed for proton beams with clinical intensities in the MedAustron synchrotron~\cite{pivi_space_2015}.

\begin{table}[thp]
    \begin{ruledtabular}
        \begin{tabular}{lc}
            \textbf{Synchrotron Setup}\\
            \colrule
            reference particle & \csix \\
            reference energy (on-momentum)& 7.00\,MeV/u \\
            nominal energy (off-momentum operation) & 6.97\,MeV/u \\
            hor./ver.\ tune $Q_x$/$Q_y$ (on-momentum) & 1.74/1.78 \\
            hor./ver.\ chrom.\ $Q'_x$/$Q'_y$ & -4/-4 \\
            \colrule
            \textbf{Injected Beam Parameters}\\
            \colrule
            inj.\ energy \hetwo & 7.03\,MeV/u \\
            inj.\ energy \csix & 6.97\,MeV/u \\
            inj.\ intensity \hetwo & $6.6\cdot10^{8}$\,ions/turn \\
            inj.\ intensity \csix & $1.2\cdot10^{8}$\,ions/turn \\
            inj.\ RMS mom. spread \hetwo/\csix & $7\cdot10^{-4}$ \\
            inj.\ RMS emitt. norm. \hetwo/\csix & $2\cdot10^{-7}$\,m\,rad \\
            inj.\ window \hetwo/\csix & 30\,µs ($\approx$\,14 turns) \\
            \colrule
            \textbf{Injection Bump}\\
            \colrule
            rise time & 120\,µs ($\approx$\,56 turns)\\
            decay time & 80\,µs ($\approx$\,38 turns)\\
        \end{tabular}
    \end{ruledtabular}
    \caption{Parameters for the simulation of the double multi-turn injection at the MedAustron synchrotron.}
    \label{tab:simulationInput}
\end{table}

Figure~\ref{fig:particleTrackingSimulations} illustrates the simulated impact of the second injection bump amplitude on the achieved helium-to-carbon intensity ratio and total injection efficiencies. A bump amplitude of 41\,mm displaces the horizontal closed orbit all the way to the injection septum blade. For the simulation parameters applied in this example, the carbon injection efficiency is maximized ($\approx 60$\,\%) for initial orbit bump amplitudes of approximately 35\,mm. Notably, varying the initial bump amplitude by $\pm5$\,mm reduces the carbon injection efficiency (Fig.~\ref{fig:particleTrackingSimulations} bottom, blue line) only down to around $50\,$\%, which could provide flexibility for tailoring the helium ion beam contribution. The simulation results further suggest, that varying the initial bump amplitude over the range of ${\pm}5\,$\,mm facilitates a configuration of the helium and carbon ion ratios between $N_\text{He}/N_\text{C}\approx0$ and ${1.6}$ at flat bottom, considering the above-cited injected beam currents. Specifically, the desired mixing ratio of 10\,\% helium and 90\,\% carbon ions would be achieved by setting the second injection bump's maximum amplitude to around 37\,mm. However, the proximity of this orbit to the septum blade not only decreases the carbon injection efficiency but also increases the sensitivity of the surviving helium content to additional errors and jitters. Consequently, when assuming similar capture, acceleration, and extraction efficiencies for both ion species and aiming for around 10\,\% helium content, it may be beneficial to lower the current of the injected helium beam. This could be accomplished by retuning the helium source for less extracted intensity or by selectively using degrader elements only during the helium injection.

However, assessing the need for such measures is subject to future developments, as the ion ratio is not only affected by the injection bump configuration but also by differences in transmission in the consequent beam transport stages, particularly the capture in the presence of the injection energy offsets between the helium and carbon ions.
Still, in this context, it is worth noting that some of these differences in transmission for helium and carbon ions may be attributed to differences in their horizontal phase space distributions after the sequential injection.

\begin{figure}[htp]
    \centering
    \includegraphics[width=\linewidth]{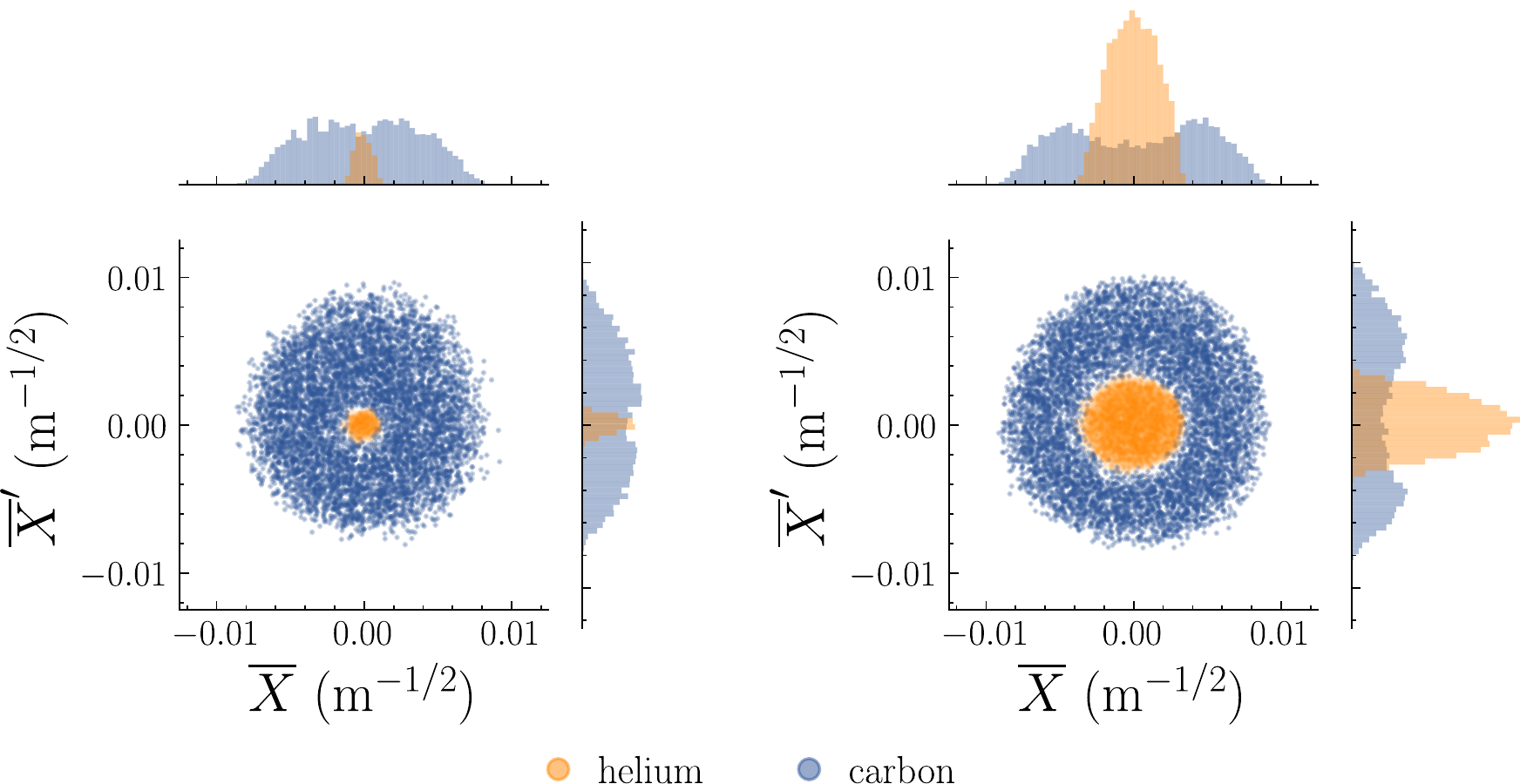}
    \caption{Simulated horizontal phase space distributions and projected profiles for around 10\,\% (left), and 50\,\% (right) helium content.}
    \label{fig:phaseSpacePlots}
    \vspace{-10 pt}
\end{figure}

Figure~\ref{fig:phaseSpacePlots} qualitatively displays the mixed beam distribution for two different amplitudes of the second injection bump, yielding 10\,\% (left) and 50\,\% helium contribution (right), respectively. The latter features a hollow horizontal carbon distribution, filled by a helium core as expected from the results of the analytical estimations shown in Fig.~\ref{fig:distributionAnalysis}. This distortion is significantly reduced for a mixed beam creation with 10\,\% helium contribution. 

The difference in the horizontal phase space distributions may have implications on the time evolution of the extracted helium-to-carbon ratio throughout the spill~\cite{renner_towards_2024}. Quantifying the extent of these implications is subject to further investigations. A brief discussion on this can be found in Section~\ref{ch:slow_extraction}.

\section{Experimental realization}
\noindent This section describes the implementation of the proposed double multi-turn injection scheme at the MedAustron accelerator facility.

\subsection{Technical implementation}
\noindent The double multi-turn injection is implemented as a special machine cycle, which allows pulsing the injector twice while keeping the synchrotron magnet fields constant at injection strength. In between the two injections, the injector needs to be reconfigured for \csix\ instead of \hetwo\ injection. At the time of writing, this reconfiguration takes around 2.9\,s, which can be attributed to reconfiguring the low-energy beam transfer line (LEBT as shown in Fig.~\ref{fig:MedAustronLayout}) to inject beam into the LINAC from another ion source. After the injection process, the synchrotron is ramped normally. This sequence of two injection pulses and one synchrotron ramp is subsequently referred to as a mixed beam cycle. At this stage, the magnetic program in the synchrotron is kept as commissioned for carbon ions~\cite{pivi_status_2019}. The fully ionized carbon ion, \csix, is considered as reference ion species.

Due to hysteresis effects, a magnetization of the current-regulated switching magnet, which selects between the respective ion sources in the LEBT, is required before each mixed beam cycle. Currently, performing this magnetization procedure adds around 15\,s of idle time after each mixed beam cycle, which is significant when compared to a conventional cycle length of up to $10$\,s. With minor changes to the switching dipole operation, the magnetization could be triggered already during the acceleration and slow extraction of the mixed beam to reduce this delay. Considering newly built facilities or upgrades to existing accelerators, the time between the injections could also be mitigated by using B-field regulation of the switching magnet. 

\subsection{Injection energy adaptation}
\label{ch:injectionEnergyAdapation}

\noindent As already mentioned in Section~\ref{ch:injectionSystem}, the commissioned helium and carbon ions feature an offset in energy per mass during injection into the synchrotron due to their different charge-to-mass-ratio when being pre-accelerated in the LINAC. This offset in injection energy was quantified to be around $\Delta (E/m)\approx0.1$\,MeV/u using time-of-flight (TOF) measurements in the injector. The TOF measurements in the MedAustron injector rely on synchronized phase probe detectors positioned downstream of the LINAC to determine the ion velocity and energy from the relative phases between the beam and the injector master oscillator at 216.816\,MHz \cite{repovz_development_2022}. For the measurements within this work, two phase probes with $7.16$\,m drift length located downstream of the debuncher cavity were used to obtain the beam energy at injection into the synchrotron.

\begin{figure}
    \centering
    \includegraphics[width=\linewidth]{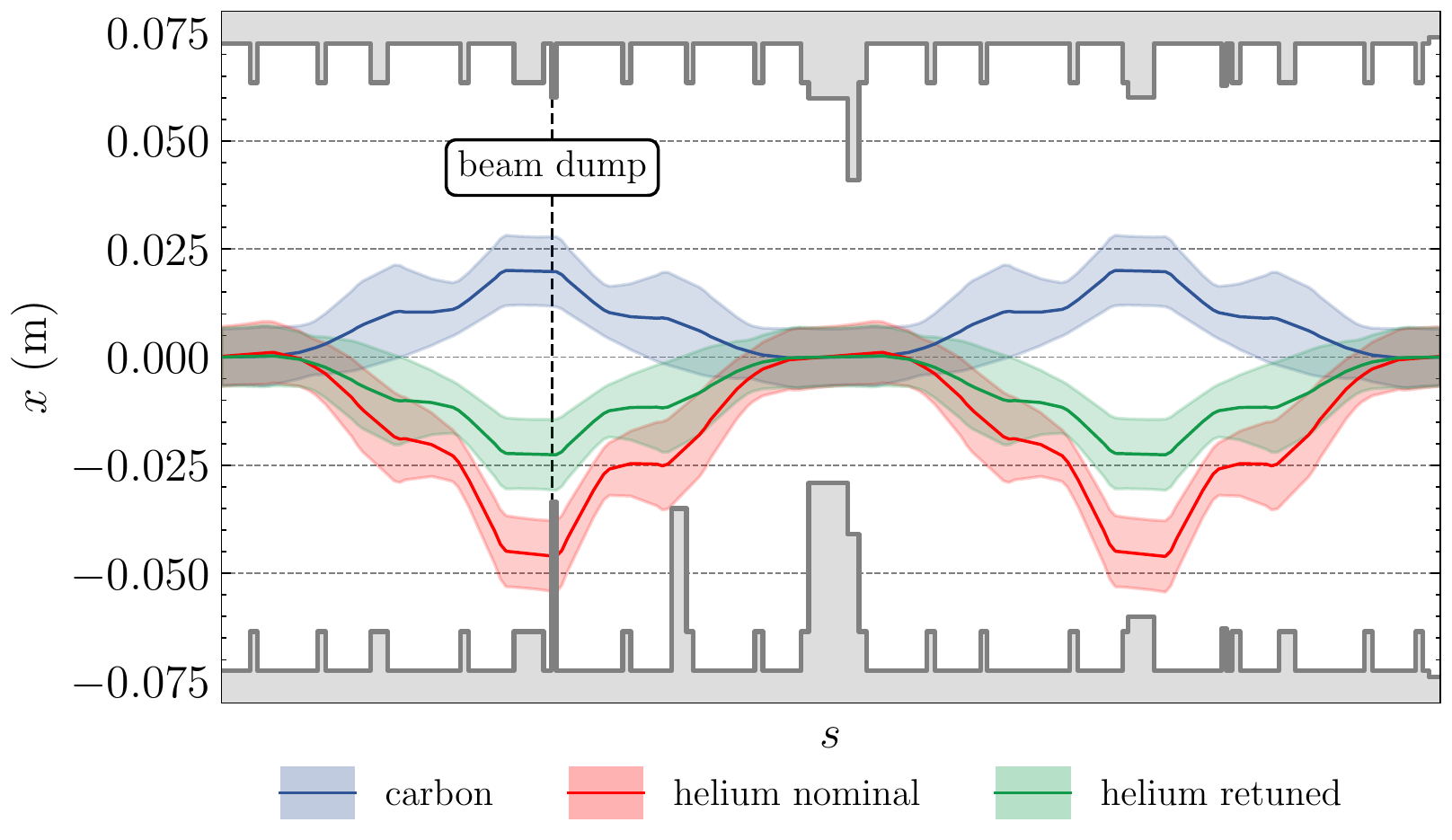}
    \caption{Dispersive beam orbits $x(s)=D_x(s)\frac{\Delta B\rho}{B\rho}$ and RMS envelopes $\sigma_x=\sqrt{\frac{\epsilon_{n,x}\beta_x(s)}{\beta\gamma}+\left(D_x(s)\sigma_{\nicefrac{\Delta B\rho}{B\rho}}\right)^2}$ of the helium beam before (\enquote{nominal}) and after energy adaptation in the injector (\enquote{retuned}) according to the parameters presented in Section~\ref{ch:injectionSystem}. The carbon beam envelope is shown for reference.}
    \label{fig:envelope_plot}
\end{figure}

Injecting the commissioned helium beam~\cite{gambino_status_2024} into the MedAustron synchrotron with magnetic field settings configured for carbon ion operation is not possible, as the associated rigidity offset of the helium beam, calculated from equation~\eqref{eq:dEE} in Appendix~\ref{ch:appendixRequirements}, would result in the beam hitting a beam dump aperture in the synchrotron as indicated by the red beam envelope in Fig.~\ref{fig:envelope_plot}.

To reduce the dispersive offset of the injected helium ions at the beam dump and thereby enable the injection of the helium ions into the synchrotron configured for carbon ions, the injection energy of \hetwo\ was adapted by retuning the injector cavities. This adjustment involved lowering the amplitude of the RF wave in the IH-mode drift tube LINAC and modifying the relative phase of the debunching cavity wave. A total energy reduction of approximately 0.04\,MeV/u to around 7.03\,MeV/u was achieved before beam losses within the LINAC were observed. The machine setting at around 7.03\,MeV/u injection energy was not fully optimized, as its performance was deemed sufficient for further investigations. As evident from the green beam envelope in Fig.~\ref{fig:envelope_plot}, the helium with reduced injection energy exhibits a smaller dispersive offset and remains within the aperture constraints at the beam dump. The retuning of the cavities, which alters the energy and momentum spread of the injected beam, in combination with steering and optics offsets introduced from the change in energy, yield an overall lower injected helium intensity, which was estimated to be around 100\,µA for the simulations in Section~\ref{ch:particleTrackingSimulation}, compared to the nominal helium machine settings, which yields around 180\,µA of \hetwo.

The injection energies for carbon and helium as commissioned and with reduced energy, as well as the reference energy of the synchrotron at injection, are summarized in Table~\ref{tab:energyAdaptation}.

\begin{table}[htp]
    \begin{ruledtabular}
        \begin{tabular}{lc}
            & \textbf{$\nicefrac{\bm{E}}{\bm{m}}$ (MeV/u)} \\
            \colrule
            \csix\ reference (on-momentum) & 7.00\\
            \csix\ nominal& 6.97$\pm$0.01\\
            \hetwo\ nominal & 7.07$\pm$0.01\\
            \hetwo\ retuned & 7.03$\pm$0.01\\
        \end{tabular}
    \end{ruledtabular}
    \caption{TOF measurement of injection energies per mass $\nicefrac{E}{m}$ of the \hetwo\ and \csix\ beams before (\enquote{nominal}) and after tuning of the LINAC and debuncher cavity (\enquote{retuned}).}
    \label{tab:energyAdaptation}
    \end{table}

\subsection{\texorpdfstring{Double multi-turn injection \\ and flat bottom measurements}{Double multi-turn injection and flat bottom measurements}}

\noindent For the presented measurements, the second injection bump maximum amplitude was set to around 32.5\,mm.
If only losses due to the double injection bump are taken into account, this would be significantly lower than the suggested setting for 10\,\% helium content derived from the particle tracking simulations in Section~\ref{ch:particleTrackingSimulation}. However, for the presented initial proof-of-principle measurements, a higher concentration of helium ions, achieved by setting a lower maximum bump amplitude, was preferred, particularly as further helium ion losses are expected during further beam transport, particularly during capture.

\begin{figure}
    \centering
    \includegraphics[width=\linewidth]{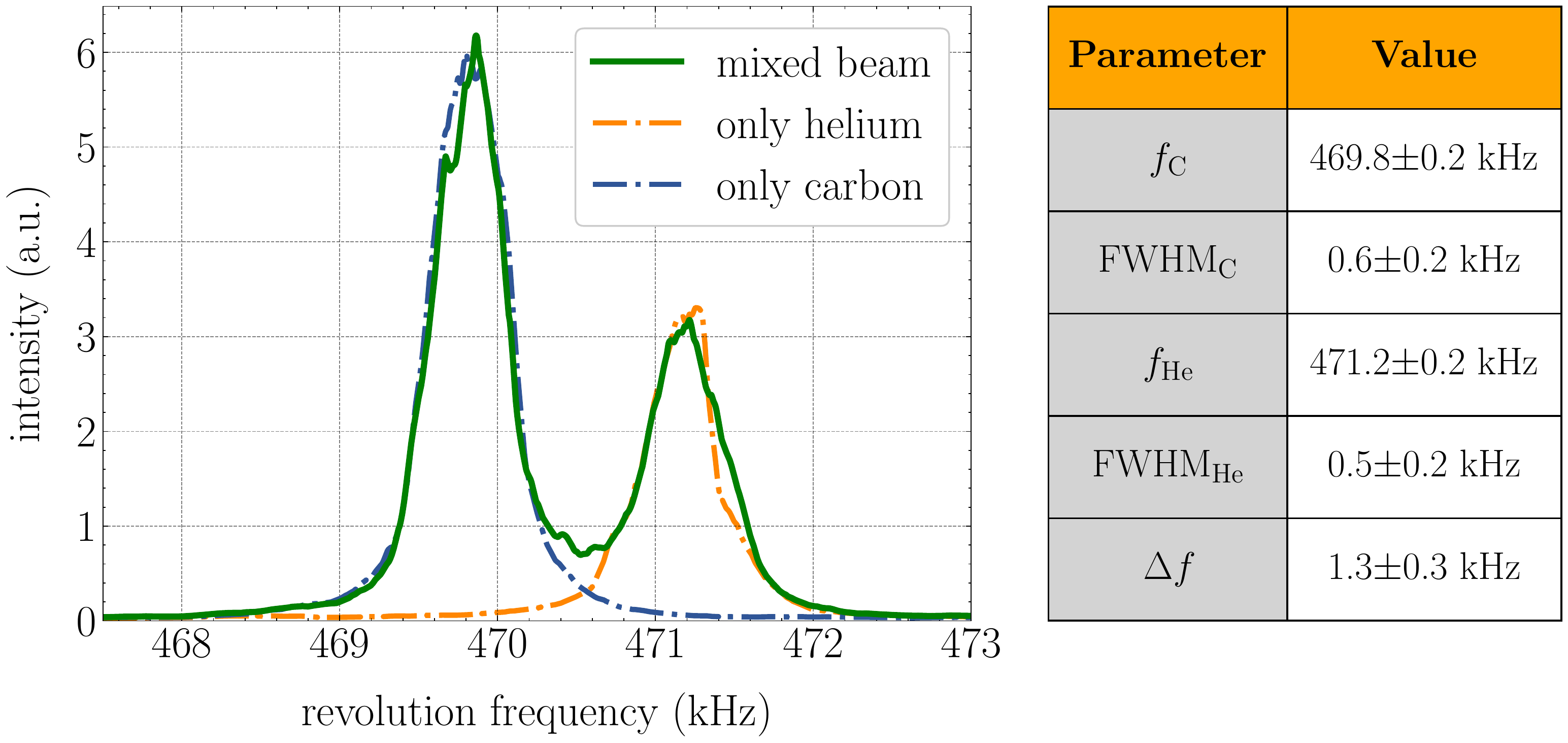}
    \caption{Revolution frequency spectrum of the coasting mixed helium and carbon beam measured after the double multi-turn injection with the center frequencies $f_\text{C/He}$ and the full width at half maximum (FWHM) frequency spreads $\text{FWHM}_\text{C/He}$.}
    \label{fig:schottkyMeasurement}
\end{figure}

After the sequential injection, the frequency spectrum of the coasting mixed beam was measured with a Schottky monitor \cite{osmic_overview_2012}. The obtained frequency spectrum, shown in Fig.~\ref{fig:schottkyMeasurement}, features a double peak structure, which is a direct consequence of the injection energy per mass offset. Performing the same injection scheme while deliberately dumping the helium or carbon ions in the injector allows for the clear identification of the two ion species. The spectra obtained from injecting only carbon or only helium coincide with the two peaks of the mixed spectrum. The frequency offset in the Schottky spectrum can be compared to the previously measured TOF injection energy per mass offset using 
\begin{equation}
    \label{eq:dEmEm_calc}
    \frac{\Delta\left(\nicefrac{E}{m}\right)}{\nicefrac{E}{m}}  = -\frac{\beta^2}{\eta}\left[\frac{\Delta f}{f} + \left(\eta+\frac{1}{\gamma^2}\right)(\frac{1}{\chi}-1)\right],
\end{equation}
which can be derived from equations~\eqref{eq:dEE} in Appendix~\ref{ch:appendixRequirements} and~\eqref{eq:appendixFrequency} in Appendix~\ref{ch:appendixFrequencyOffset}. The slippage factor at injection energy is $\eta\approx-0.74$, $\beta$ and $\gamma$ are the relativistic parameters for carbon ions at the reference energy, and $\chi$ is the ratio of charge-to-mass ratios as defined in equation~\eqref{eq:chi} in Appendix~\ref{ch:appendixRequirements}. The relative offset of the ions from the reference energy per mass, as obtained from equation~\eqref{eq:dEmEm_calc}, can then be used to determine the total injection energy per mass offset between the helium and carbon ions.
\begin{equation}
    \label{eq:explicidDeltaEm}
    \Delta\left(\frac{E}{m}\right) = \left[\left(\frac{\Delta(\nicefrac{E}{m})}{\nicefrac{E}{m}}\right)_\text{He}-\left(\frac{\Delta(\nicefrac{E}{m})}{\nicefrac{E}{m}}\right)_\text{C}\right]\left(\frac{E}{m}\right)_\text{ref}
\end{equation}

Table~\ref{tab:energyOffsetSchottky} shows that the energy per mass offsets computed from the acquired frequency spectrum and equations~\eqref{eq:dEmEm_calc} and \eqref{eq:explicidDeltaEm} agree with the injection energy offsets obtained from the TOF measurement within the error margins, which are calculated from the indicated errors in Table~\ref{tab:energyAdaptation} and Fig.~\ref{fig:schottkyMeasurement}.

\begin{table}[htp]
    \begin{ruledtabular}
        \begin{tabular}{lc}
            \textbf{meas. method} & \textbf{$\Delta(\nicefrac{\bm{E}}{\bm{m}})$}\\
            \colrule
            TOF& 0.06$\pm$0.01 \\
            Schottky& 0.05$\pm$0.01
        \end{tabular}
    \end{ruledtabular}
    \caption{Measured offset in injection energy per mass $\Delta(\nicefrac{E}{m})$ via TOF in the injector and Schottky measurements in the synchrotron. }
    \label{tab:energyOffsetSchottky}
\end{table}

\subsection{Capture and acceleration}
\label{ch:capture_acceleration}

\noindent For the presented measurements, the capture and acceleration to 262.3\,MeV/u was performed using the machine settings of the commissioned carbon beam. As these settings, and in particular, the fixed capture frequency, are optimized for the carbon injection energy~\cite{pivi_status_2019}, the capture efficiency for the helium ions with the observed revolution frequency offset is expected to be low. The flat top beam current after the acceleration ramp is around 300\,µA, which corresponds to around $10^8$ to $4\cdot10^{8}$ ions in the synchrotron. The exact value depends on the unknown mixing ratio after the acceleration. Optimizations to the capture and acceleration process, such as the reduction of the described energy per mass offset of the injected beams as well as quantification concepts for the beam composition are subject to further developments. 

It is worth noting that the helium and carbon ions captured and accelerated with the single-harmonic RF system at MedAustron feature slightly different rigidities after acceleration according to equation~\ref{eq:appendix_rigidity_offset} in Appendix~\ref{ch:appendixFrequencyOffset}.

\subsection{Slow extraction}
\label{ch:slow_extraction}
\noindent Slow extracting this ion mix for treatment monitoring requires not only a constant total extraction rate but ideally also an approximately constant helium-to-carbon ratio throughout the entire spill. We aim for variations within approximately 5\,\%, although this target is currently not derived from any dedicated imaging system or treatment planning strategy and will be updated as more detailed input becomes available. Achieving this level of stability is complicated by slight differences in the extraction dynamics of the two ion species due to two main effects~\cite{renner_towards_2024}.

The first is the total rigidity offset (see equation~\ref{eq:appendix_rigidity_offset} in Appendix~\ref{ch:appendixFrequencyOffset}), which, in the presence of non-zero horizontal chromaticity $Q'_x\neq0$, induces a horizontal tune shift ${\Delta Q_x = Q'_x \cdot \frac{\Delta \left(B\rho\right)}{\left(B\rho\right)}}$ resulting in different tune distances from the resonance and, consequently, different stable phase space areas for the two ion species.
In addition to this effect, which is also present for mixed beams generated in a single ion source, the double-injection scheme causes differences in the horizontal phase-space distributions (see Section~\ref{ch:particleTrackingSimulation}). Hence, helium and carbon ions can occupy different portions of the stable phase space region prior to extraction.

While a detailed investigation of these effects and their implications is left for future studies, the presented proof-of-principle experiment was limited to demonstrating the extraction of both ion species, independent of the time structure.

For the presented measurements, the mixed beam was extracted using a non-optimized RF phase displacement extraction \cite{renner_investigating_2024, kuhteubl_slow_2024} as incompatibilities in the control system implementation of the betatron core with the double injection scheme prohibited extracting the mixed beam via the betatron core extraction method, which is conventionally used for irradiation at MedAustron. At the time of the experiments, constraints in the RF system implementation required that during the phase displacement extraction the RF frequency sweeps needed to be set up to extract the entire beam in three sweeps of around \SI{100}{\micro\second} each instead of a continuous spill. During each sweep, approximately one third of the ions in the synchrotron, i.e.\ 100\,µA or $3\cdot10^7$ to $10^8$ ions, are extracted. The extracted beam was steered into the non-clinical irradiation room (IR1 as shown in Fig.~\ref{fig:MedAustronLayout}) using the commissioned carbon machine settings.

Future work will explore RF knockout (RFKO) for extracting the ion mixture, which can be experimentally employed at MedAustron~\cite{kuehteubl_investigating_2023,kuhteubl_slow_2024, renner_investigating_2024}. Applying different RFKO excitation signals with different central frequencies and amplitude modulations may facilitate more precise regulation of the helium-to-carbon ratio over time~\cite{renner_towards_2024}.

\subsection{\texorpdfstring{Detection in the \\ irradiation room}{Detection in the irradiation room}}
\label{ch:Detection_Irradiation_Room}

\begin{figure}
    \centering
    \includegraphics[width=\linewidth]{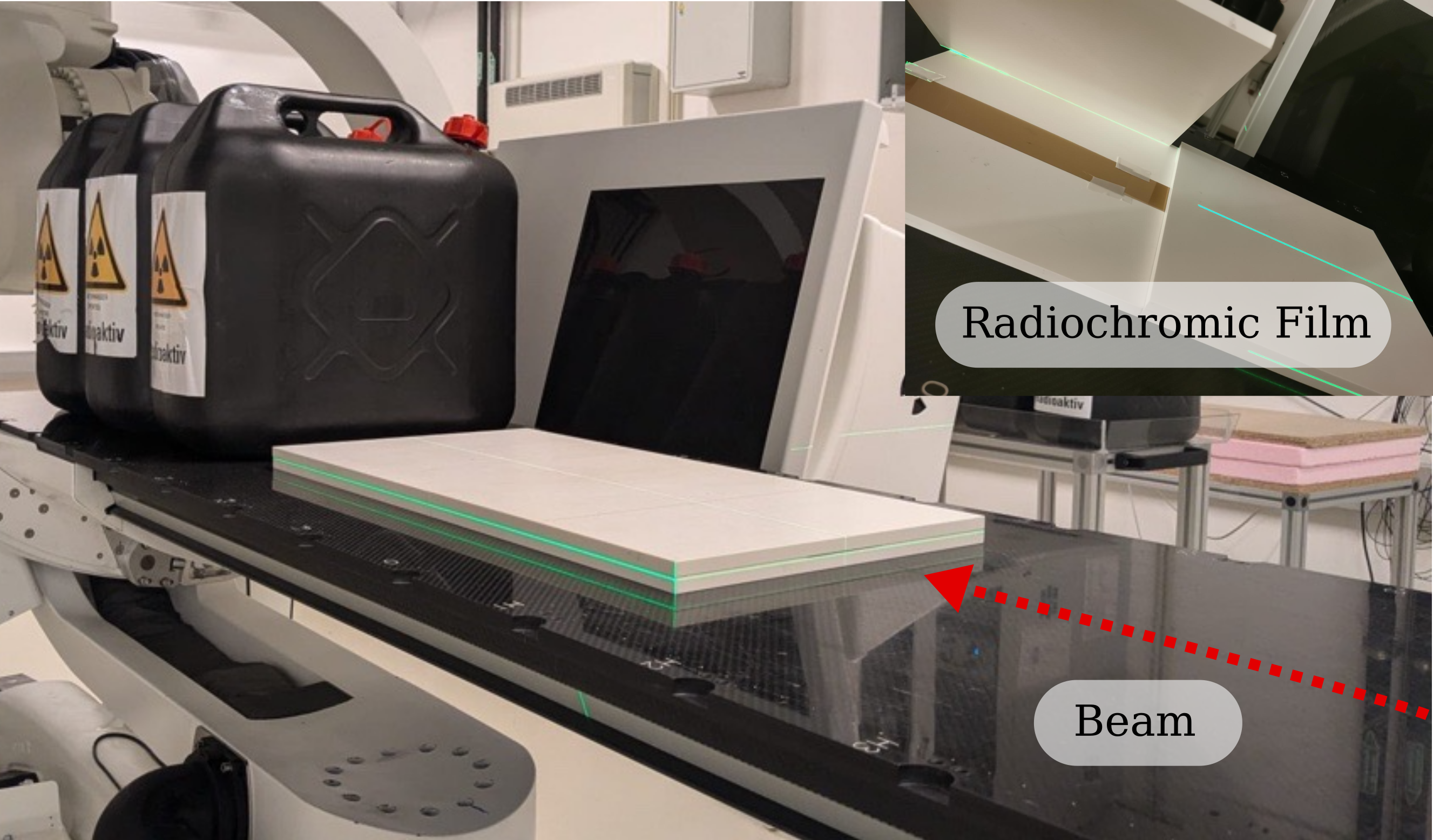}
    \caption{Experimental setup of the mixed beam detection with radiochromic films in the non-clinical irradiation room (IR1) at MedAustron.}
    \label{fig:experimentalSetup}
\end{figure}

\begin{figure*}
    \centering
    \includegraphics[width=\linewidth]{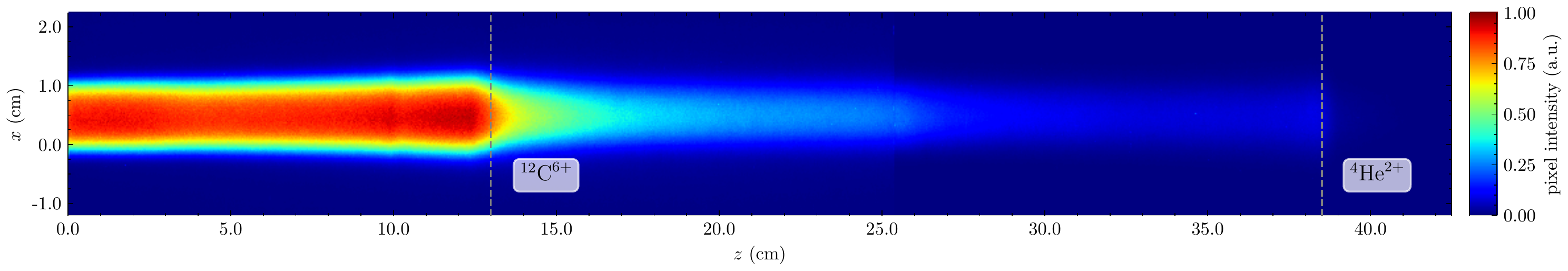}
    \caption{
    Radiochromic film measurement of 30 mixed helium and carbon ion beam spills with indications for the measured R80 ranges. Due to the higher dose deposition from carbon ions, the film saturates before the helium Bragg peak becomes visible. As a result, this measurement serves only as a qualitative demonstration of the simultaneous delivery of both ion species.}
    \label{fig:filmAnalysis}
\end{figure*}

\noindent The first detection of the mixed beam in IR1 was achieved via a radiochromic film measurement. To cover the full range of the beam, two sheets of GAFChromic EBT3 film \cite{noauthor_ashland_2024} were stitched together and placed between two 1\,cm PTW RW3 slabs with water equivalent thickness of 1.04\,cm aligned along the $xz$-plane at the isocenter. The water equivalent thickness of the RW3 slabs had been measured at MedAustron for different research projects and is available internally. A picture of the experimental setup is shown in Fig.~\ref{fig:experimentalSetup}.

In total, 30 mixed beam spills were applied to the measurement setup. The pixel intensity on the irradiated film is shown in Fig.~\ref{fig:filmAnalysis}. The Bragg peaks of \csix\ and \hetwo\ are distinguishable and their measured R80 ranges (12.99\,cm for \csix\ and 38.51\,cm for \hetwo), which are retrieved from fitting an asymmetric Lorentzian and a Gaussian convolved with an error function to the pixel intensity profile of the carbon and helium peak. These functions were chosen as they replicate the curve shapes of the projected intensity profile to retrieve the R80 ranges. As evident from the estimated R80 ranges and Fig.~\ref{fig:filmAnalysis}, the helium ions exhibit an approximately three times longer range than carbon ions.

A common steering offset of around 0.4\,cm to the isocenter is present, which can be attributed to the non-optimized extraction setup. Below 10\,cm range, several artifacts are apparent. The beam envelope is not constant but decreases, reaching a minimum at around 5\,cm range. Additionally, a local intensity maximum is observed at around 10\,cm. While these artifacts usually suggest contamination with heavier ion species, in this case, the measurement is likely influenced by scattering from the positioning bench visible in the experimental setup in Fig.~\ref{fig:experimentalSetup}, and beam spot drifts during the extraction.

Overall, the presented film measurement confirms the presence of \csix\ and \hetwo\ ions in the beam. However, it should be noted that the measurement is subject to errors, including artifacts and uncertainties in the acquired R80 ranges arising from the film stitching, saturation effects, and reliance on pixel intensity rather than the actual dose.

\begin{figure}[tp]
    \centering
    \includegraphics[width=\linewidth]{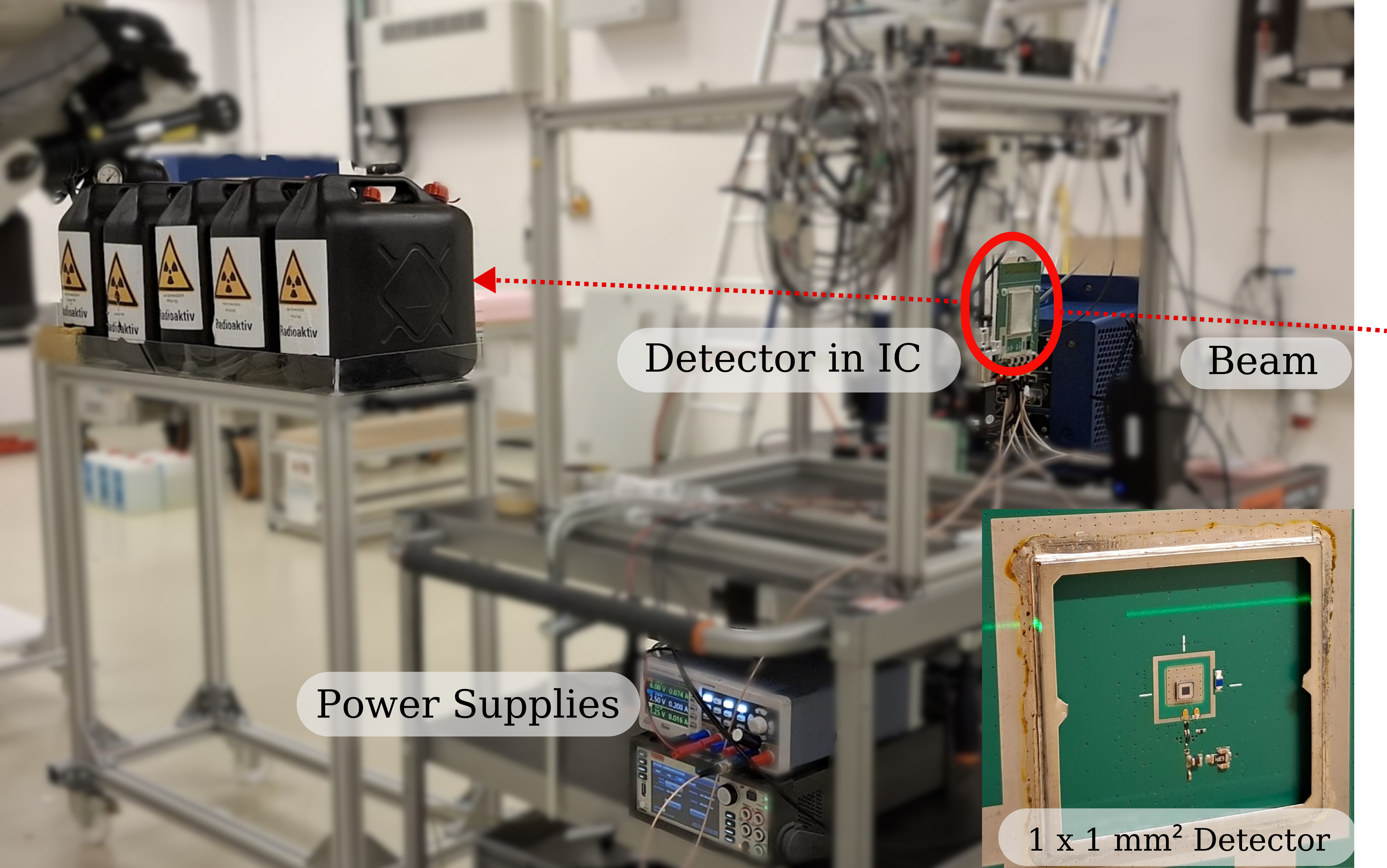}
    \caption{Experimental setup of the LET-based \hetwo\ and \csix\ discrimination via a silicon LGAD. A close-up of the detector is shown in the lower right corner.}
    \label{fig:dEdxMeasurement}
\end{figure}

A complementary measurement was performed by measuring the energy deposition of the extracted ions in a silicon detector. At 262.3\,MeV/u \hetwo\ deposits around \SI[per-mode=symbol]{2.8}{\kilo\electronvolt\per\micro\meter}, whereas \csix\ deposits about \SI[per-mode=symbol]{25.1}{\kilo\electronvolt\per\micro\meter}~\cite{ziegler_srim_2010}. This difference in energy deposition per path length, also called linear energy transfer (LET), allows to distinguish the particles via the detector signal.
The experimental setup consisted of a \SI{50}{\micro\meter} thick silicon (LGAD) detector from the FBK UFSD2 production with 1\,mm$\times$1\,mm active area~\cite{sola_first_2019}, mounted on an Infineon BFR840-based transimpedance amplifier readout-board.
LGAD detectors allow for fast signals ($< \SI{2}{\nano\second}$ pulse length) proportional to the deposited energy while providing a high signal-to-noise ratio due to their internal amplification. 
For the detection of the mixed beam, the LGAD was operated with a reverse bias of \SI{90}{\volt}. The signals from the front-end electronics were further amplified by a \SI{22}{\decibel} Minicircuits ZX60-14LN-S+ broad-band amplifier.
To digitize the waveforms, a Rohde \& Schwarz RTO6 oscilloscope was used. It was configured to the \SI{12}{\bit} HD Mode with a limited analog bandwidth of \SI{1}{\giga\hertz} for the best noise performance.
The oscilloscope readout was triggered directly by a trigger on the rising edge of the waveform.
Figure~\ref{fig:dEdxMeasurement} shows a picture of the experimental setup mounted at the isocenter (IC) of IR1.

\begin{figure}[htp]
\vspace{5pt}
    \centering
    \includegraphics[width=\linewidth]{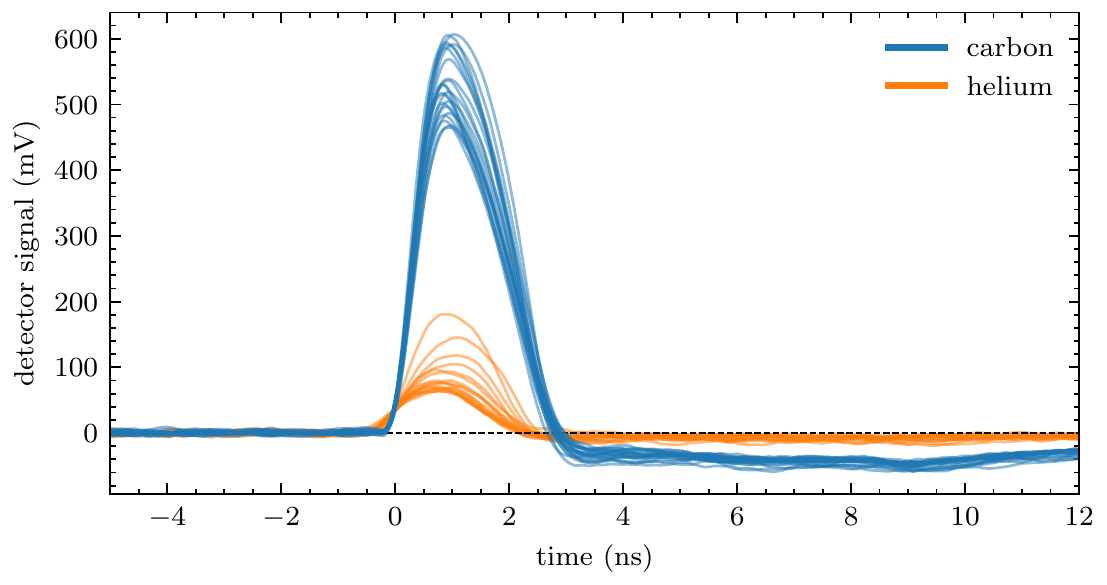}
    \caption{Typical measured detector waveforms for 262.3\,MeV/u \hetwo\ and \csix\ ions. For each ion species, 20 events are shown.}
    \label{fig:dEdxWaveforms}
\end{figure}

Figure~\ref{fig:dEdxWaveforms} shows the typical detector waveforms for \hetwo\ and \csix\ ions.
Beams containing only one ion species were obtained by inserting the Faraday cup in the source branch of the other ion species to deliberately inhibit its injection.
Due to the large difference in LET between the \hetwo\ and \csix\ ions, the ion species can be clearly distinguished by the signal's amplitude. As the front end of the detector readout is a transimpedance amplifier, the signal is proportional to the detector current, and the integral thereof to the deposited energy. The negative signal tails for the carbon hits are a result of a signal overshoot in the readout electronics. This overshoot is also visible for the helium signals, albeit much smaller.

\begin{figure}[t]
    \centering
    \includegraphics[width=\linewidth]{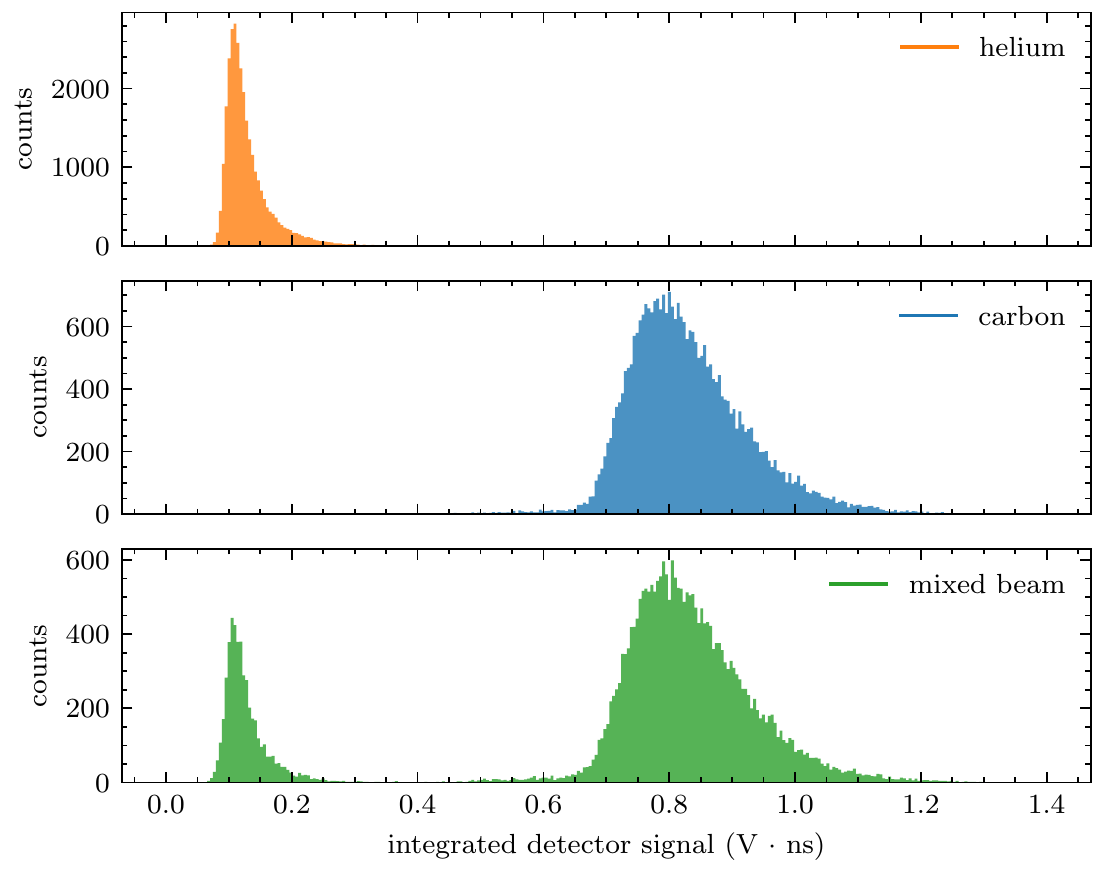}
    \caption{Histograms of the integrated detector signal for \num{30000} events at the beginning of four consecutive slow extractions in helium-only, carbon-only, and mixed beam conditions.}
    \label{fig:dEdxHisto} 
\end{figure}

Figure~\ref{fig:dEdxHisto} shows the histograms of the integrated detector signal for only helium ions, only carbon ions, and the mixed beam. The integrated area is computed taking into account only the positive part of the waveform above a fixed threshold. This threshold is chosen as the $1\sigma$ standard deviation calculated from the first nanosecond of the acquisition before the ion hit.
The histograms featuring only a single ion species show the expected Landau-like energy loss distributions. The distributions of \hetwo\ and \csix\ are well separated.
For the mixed beam, both distributions, coinciding with \hetwo\ and \csix\ irradiation, are present.
Setting a threshold in the middle of the two distributions, e.g.\ 0.4\,Vns in Fig.~\ref{fig:dEdxHisto}, around \num{5000} \hetwo\ and \num{25000} \csix\ ions are counted, yielding a mixing ratio of around 1:5 during the acquisition window. This demonstrates the possibility of performing particle-by-particle discrimination between \hetwo\ and \csix\ due to their mass (and LET) difference and confirms again the simultaneous presence of helium and carbon ions in the extracted beam. 


In these measurements, it was only possible to acquire the first \num{7500} trigger events, corresponding to a few milliseconds of acquisition time. This limitation is enhanced by the chosen configuration of the phase-displacement slow extraction and the resulting spill structure consisting of three fast 100\,µs extraction sweeps leading to high beam intensities of approximately $3\cdot10^7$ to $10^8$ ions/100\,µs depending on the unknown ion mixing ratio after acceleration (see Section~\ref{ch:slow_extraction}) and pile-up in the detector. Moreover, the active detector area of 1\,mm$\times$1\,mm does not cover the full beam spot. Consequently, no quantification of the helium-to-carbon ion ratio throughout the entire spill can be provided with this measurement. For future measurements aiming at quantifying the helium-to-carbon ratio throughout the entire spill, the pile-up can be mitigated by improving the time structure of the slow extraction. Using a two-dimensional pixel or strip array of these LET-based detectors could allow for a time- and space-resolved determination of the \hetwo\ and \csix\ mixing ratio over the full beam spot throughout the entire spill.
To investigate the micro-spill structure, i.e.\ the high frequency component of the extracted intensity, a DC-coupled transimpedance amplifier with a bandwidth of \SI{20}{\mega\hertz} could be used, as was demonstrated for ultra-high dose rate proton beams in~\cite{waid_pulsed_2024}. In the lower frequency regime in the order of kHz, the extracted beam composition can be determined by an ionization chamber range telescope as described in~\cite{kausel_recent_2025}.

\section{Discussion}
\noindent While the implementation of the sequential injection poses several challenges, it also offers distinct advantages compared to a mixed beam generation in a single ion source.

One of the challenges associated with generating the beam using the double-injection setup, compared to employing a single ion source, arises from differences in the horizontal and longitudinal phase space distributions of helium and carbon ions after the sequential injection. The energy per mass offset of the two ion species after injection will need to be reduced to ensure a sufficient capture efficiency for both ion species and avoid distortion of the RF feedback loop regulations during capture. Further, the horizontal phase space features helium in the core enclosed by the hollow carbon ion distribution. Possible implications for the time evolution of the extracted helium-to-carbon ratio throughout the slow extraction are subject to further investigations.

Whereas the beam dynamics challenges described above are expected to be solvable, another major obstacle for potential future applications in ion beam radiotherapy lies in the efficiency of the double injection setup. Although the time between the two injections can be substantially reduced with minor upgrades in hardware and operational procedure, the double pulsing of the injector will always take more power and time compared to a mixed beam generation in a single ion source. After the initial research phase, the different production schemes will need to be evaluated in terms of sustainability and economic efficiency. 

However, despite these challenges, the sequential injection scheme is worth studying as it features certain advantages over creating the mixed beam within a single ion source.
Firstly, the presented results indicate that the double multi-turn injection scheme allows for the delivery of mixed helium and carbon ion beams at state-of-the-art medical carbon and helium synchrotron facilities without requiring significant modifications to the existing infrastructure.

Another advantage is the inherent purity of the ion mix. Generating a mixed beam within a single ion source can lead to contamination through elements introduced by air leaks or surface outgassing of vacuum elements. Especially the combination of \heone\ and \cthree\ is prone to contamination with \ofour as these ion species can be easily produced by ECR ion sources. This issue was observed in the investigations at GSI in late 2023~\cite{galonska_first_2024}. The risk of oxygen or nitrogen contamination is expected to be reduced when generating a mixed \hetwo\ and \csix\ beam in a single ion source as both \nseven\ and \oeight\ exhibit significantly larger ionization energies compared to \hetwo\ and \csix. When generating the ion mix by the sequential injection scheme, the risk of contamination is assumed to be even smaller as it consists of clinically commissioned helium and carbon research and treatment beams. The carbon beam, which consists of \cfour\ ions before being stripped to \csix after the LINAC, can be produced as a pure beam as no charge state of oxygen or nitrogen overlaps with \cfour. For the generation of \hetwo\ beams the probability of producing \oeight\ or \nseven\ is very small as the ionization energy of \hetwo\ is significantly smaller compared to \oeight\ or \nseven.

Further, in facilities with RF systems that support the application of multiple independent RF frequencies, or facilities that enable bunch-to-bucket injection, a sequential injection of the two ion species may facilitate separating helium and carbon longitudinally. The longitudinal separation might enable advanced beam manipulations in preparation for or during the extraction process, which in turn could allow for customizing the timing structure of the delivered mixed beam. If any future facilities are planned to produce mixed beams via sequential injection, it could also be worth exploring the option of positioning the injection septum in a dispersive region. When combined with an energy-ramping cavity in the LINAC, this setup may allow for longitudinal painting during injection instead of the conventional horizontal painting, as done in the injection system of LEIR at CERN~\cite{benedikt_lhc_2004}. This approach could provide an additional degree of freedom for the double injection scheme. However, the related increase in energy spread would need to be assessed, particularly in the absence of beam cooling mechanisms, such as the electron cooling system installed at LEIR.

Regardless of the specific generation scheme, the characterization of the mixed beam at delivery is instrumental to future imaging research efforts. Especially, the time-resolved quantification of the ion mixing ratio over the full spill is a necessary prerequisite. As elaborated in Section~\ref{ch:Detection_Irradiation_Room}, a range telescope consisting of standard ionization chambers can be used to resolve the extracted intensity and mixing ratio with time resolution in the order of tens of milliseconds. As the presented LET-based measurement procedure relies on single particle detection, it could be used to resolve the extracted intensity and mixing ratio on much smaller time scales.

Lastly, the translation of mixed beams into a clinical tool for treatment monitoring will require further research. This includes, for example, the development of a viable imaging system capable of determining the residual helium energy over a wide range, as well as establishing robust treatment planning concepts~\cite{hardt_potential_2024, hardt_helium_2025}. In general, mixed beam treatment plans not only need to account for the additional dose deposited by the helium ions but must also consider that the helium, which features around three times longer range in matter compared the carbon ions, has sufficient range for fully traversing the patient. A conceivable approach could be to combine mono-isotopic carbon beam treatment with the mixed beam for only selected spots or energy layers with suitable helium ranges. This combination of mono-isotopic and mixed beam irradiation within a single treatment plan requires fast switching between the mixed beam and a pure carbon ion beam. Such switching is facilitated more easily with the double injection scheme compared to the generation within a single ion source, as with the sequential injection scheme the injection of helium ions can be inhibited on a spill-to-spill basis.

\section{Conclusion}
\noindent This work reports the first successful mixed \hetwo\ and \csix\ beam delivery in a medical synchrotron facility. The mixed beam was generated at the MedAustron accelerator by implementing a novel double multi-turn injection scheme, which injects the helium and carbon ions sequentially into the synchrotron. After acceleration to 262.3\,MeV/u, the mixed beam was extracted and detected in the non-clinical irradiation room using a radiochromic film and a silicon LGAD detector. The successful delivery at MedAustron indicates that the generation of mixed helium and carbon ion beams might also be possible at other state-of-the-art medical synchrotron facilities. The experimental results are supported by a detailed description of the sequential multi-turn injection process and particle tracking simulations performed for the MedAustron synchrotron. While it was shown that the proposed injection scheme achieves the mixing and allows for further acceleration and extraction of helium and carbon beams, optimizations focusing on the mitigation of injection errors, injection energy offsets, capture efficiency, ion mixing ratio throughout the spill, and overall intensity increases are necessary before the mixed beam can be applied for research purposes.

\begin{acknowledgments}
    \noindent The authors wish to express their acknowledgment to the non-clinical research partners of MedAustron and the TBU and OPS departments of the MedAustron research and treatment center for their continuous support. The authors would like to particularly thank Fabien Plassard, Mauro Pivi, Valeria Rizzoglio, and Ivan Strasik for valuable input and interesting discussions. Furthermore, the authors want to thank INFN Torino for providing the UFSD2 LGAD sensors used to detect the mixed beam in the irradiation room. The financial support of the Austrian Ministry of Education, Science, and Research is gratefully acknowledged for providing beam time and research infrastructure at MedAustron. The research underlying this work was funded as part of the RTI-Strategy Lower Austria 2027 and by the Austrian Science Fund (FWF) Erwin-Schrödinger Grant J 4762-N. The measurements with the UFSD2 LGAD have been financially supported by the Austrian Research Promotion Agency FFG, grant number 883652.
\end{acknowledgments}

\appendix

\section{\texorpdfstring{Requirements for simultaneous transport and acceleration of multi-isotopic beams}{Requirements for simultaneous transport and acceleration of multi-isotopic beams}}
\label{ch:appendixRequirements}
\noindent A general requirement for successful transport of a multi-isotopic beam is that all constituents exhibit a dispersive offset $\Delta x_D(s)$ which is considerably smaller than the size of the beam pipe aperture $A_x(s)$ at the observed location~$s$.
\begin{equation}
    \label{eq:dispersionCondition}
    \Delta x_D(s) = D(s) \cdot \frac{\Delta(B\rho)}{B\rho} < A_x(s)
\end{equation}
Here, $D(s)$ is the dispersion function, and $\Delta(B\rho)/(B\rho)$ is the relative offset in magnetic rigidity for a multi-isotopic beam. Due to the limited apertures in synchrotrons, $\Delta(B\rho)/(B\rho) \ll 1$ is required. Moreover, for successful capture and acceleration in a synchrotron with one single-harmonic RF system, the revolution frequencies of the constituents need to be similar. This implies similar velocities and consequently $\Delta(\beta\gamma)/(\beta\gamma) \ll 1$. Considering the definition of magnetic rigidity, it is immediately evident that only ions with similar mass-to-charge ratios can fulfill both constraints.
\begin{eqnarray}
    B\rho = \frac{p}{q} &=& \frac{m}{q} \beta\gamma c \nonumber\\ 
    \frac{\Delta(B\rho)}{B\rho}, \frac{\Delta(\beta\gamma)}{\beta\gamma}\ll1 &\Rightarrow& \frac{\Delta(\nicefrac{m}{q})}{\nicefrac{m}{q}} \ll 1
\end{eqnarray}

For multi-isotopic beams, the magnetic rigidity can be approximated via the logarithmic derivative~\cite{bryant_principles_1993}
\begin{equation}
    \label{eq:dBrhoBrho}
    \frac{\Delta(B\rho)}{B\rho}=\frac{\Delta\left(\nicefrac{p}{q}\right)}{\nicefrac{p}{q}} \approx \frac{\Delta\left(\nicefrac{m}{q}\right)}{\nicefrac{m}{q}} + \frac{\Delta(\beta \gamma)}{\beta \gamma}.
\end{equation}
Here, the offset in mass-to-charge ratio is given by
\begin{equation}
    \label{eq:chi}
    \frac{\Delta\left(\nicefrac{m}{q}\right)}{\nicefrac{m}{q}} = \frac{\nicefrac{m'}{q'} - \nicefrac{m}{q}}{\nicefrac{m}{q}} = \frac{1}{\chi} - 1, \ \ \ \chi = \frac{\nicefrac{q'}{m'}}{\nicefrac{q}{m}},
\end{equation}
where $m',q'$ are the mass and charge of the secondary and $m,q$ are the mass and charge of the reference particle species. The parameter $\chi$ is the ratio of the respective charge-to-mass ratios. It is chosen following the notation of multi-isotopic beams in the \textit{Xsuite} simulation framework~\cite{iadarola_xsuite_2024}.

The magnetic rigidity offset of a multi-isotopic beam can be related to the energy per mass offset by considering the relativistic energy-momentum relation
\begin{equation}
    \label{eq:energyMomentumRelation}
    \frac{E^2}{m^2} = \frac{p^2c^2}{m^2} + c^4 = \left[(\beta\gamma)^2+1 \right] \, c^4.
\end{equation}
Differentiation and elementary transformation yield
\begin{equation}
    \frac{\Delta\left(\nicefrac{E}{m}\right)}{\nicefrac{E}{m}} \approx \beta^2\frac{\Delta(\beta\gamma)}{\beta\gamma}.
\end{equation}
The relative offset in $\beta\gamma$ can be expressed in terms of the magnetic rigidity through equation \eqref{eq:dBrhoBrho} and \eqref{eq:chi}, yielding an expression for the magnetic rigidity in dependence on the relative energy per mass offset $\Delta \left(\nicefrac{E}{m}\right)/\left(\nicefrac{E}{m}\right)$ and the ratio of the charge-to-mass ratios $\chi$,
\begin{equation}
    \label{eq:dEE}
    \frac{\Delta(B\rho)}{B\rho} = \frac{1}{\beta^2} \frac{\Delta\left(\nicefrac{E}{m}\right)}{\nicefrac{E}{m}} + \frac{1}{\chi} - 1.
\end{equation}
The first term describes the relative rigidity offset caused by the relative energy per mass offset. The second term describes the additional contribution from the non-nominal charge-to-mass ratio.

\section{\texorpdfstring{Revolution frequency offset of coasting multi-isotopic beams}{Revolution frequency offset of coasting multi-isotopic beams}}
\label{ch:appendixFrequencyOffset}

\noindent As derived in detail in~\cite{bryant_principles_1993}, the contribution of the charge-to-mass ratio offset to the relative offset in magnetic rigidity has to be taken into account when computing the relative offset in revolution frequency $f=\beta c/{C}$, with $C$ being the reference path length. For $\beta\not\approx1$, ions with $\chi\neq1$ will exhibit a different velocity compared to particles with the same magnetic rigidity but $\chi=1$. According to
\begin{equation}
\label{eq:dCC}
    \frac{\Delta C}{C} = \alpha_C \frac{\Delta(B\rho)}{B\rho},
\end{equation}
with $\alpha_C$ being the momentum compaction factor, particles with the same relative magnetic rigidity offset but different charge-to-mass ratios will follow identical trajectories at different velocities and hence revolution frequencies $f$.

By transforming equation \eqref{eq:dBrhoBrho} using the relation 
\begin{equation}
    \frac{\Delta(\beta\gamma)}{\beta\gamma} \approx \gamma^2 \frac{\Delta\beta}{\beta}
\end{equation}
the relative offset in velocity can be expressed as
\begin{equation}
    \frac{\Delta\beta}{\beta} = \frac{1}{\gamma^2}\left[\frac{\Delta(B\rho)}{B\rho}-\left(\frac{1}{\chi}-1\right)\right].
    \vspace{5pt}
\end{equation}
Substituting the relative path length offset from equation~\eqref{eq:dCC} into the logarithmic derivative of the revolution frequency
\begin{equation}
    \label{eq:logDerivf}
    \frac{\Delta f}{f} \approx \frac{\Delta\beta}{\beta} - \frac{\Delta C}{C}
\end{equation}
yields 
\begin{equation}
    \label{eq:appendixFrequency}
    \frac{\Delta f}{f} = -\eta\frac{\Delta(B\rho)}{B\rho} -\frac{1}{\gamma^2}\left(\frac{1}{\chi}-1\right),
\end{equation}
where $\eta=\alpha_C-\gamma^{-2}$ is the slippage factor. This relation features two contributions. The first term expresses the relative frequency offset proportional to the relative magnetic rigidity offset, as also present for mono-isotopic beams. The second term can be interpreted as a correction term, which considers that the rigidity offset not only originates in velocity but also in the offset of charge-to-mass ratios.

When being captured or accelerated in a single-harmonic RF system with harmonic number $h=1$, the revolution frequencies of the ions must match the RF frequency $\Delta f / f = 0$. According to equation~\ref{eq:appendixFrequency} this leads to a rigidity offset
\begin{equation}
    \frac{\Delta (B\rho)}{B\rho} = -\frac{1}{\eta\gamma^2}\left(\frac{1}{\chi}-1\right).
    \label{eq:appendix_rigidity_offset}
\end{equation}

\section{\texorpdfstring{Supplementary materials to the horizontal acceptance analysis}{Supplementary materials to the horizontal acceptance analysis}}
\label{ch:appendixDetailsAnalysisHorizontalPhaseSpace}
For each turn $i$ during the second (carbon) injection bump, the condition for particle preservation in normalized phase space $(\overline{X}, \overline{X}')$ is given by requiring that $\overline{X}<\overline{X}_S$, where $\overline{X}$ is the normalized particle position and $\overline{X}_S$ is the normalized injection septum aperture. Considering that the distance from the beam center to the aperture changes during the injection bump rise and decay, it is convenient to apply a coordinate transformation that shifts the time dependency to the constraints and ensures that the beam is centered in the origin of the normalized phase space at all times. The transformed constraints $ \Delta\overline{\bm{X}}_i$ for any turn~$i$ then correspond to the momentary distance of the beam center $\overline{X}_{i}$ to the injection septum $\overline{X}_S$
\begin{equation}
    \Delta \overline{\bm{X}}_i = (\overline{X}_S - \overline{X}_{i}, \overline{X}'_i)^T,
\end{equation}
where $\overline{X}'_i$ is the normalized angle with arbitrary value. Successively transforming $\Delta \overline{\bm{X}}_i$ to the observation turn~$j$ via the one-turn map $\bm{M}$ yields the parametric form of the constraint.
\begin{equation}
    \overline{\bm{X}}_{i,j} = \bm{M}^{j-i} \Delta \overline{\bm{X}}_i 
\end{equation}
This expression allows to plot all constraints within one phase space diagram at turn $j$.

\bibliography{references.bib}

\end{document}